\pdfoutput=1
 \documentclass[prd, 12pt,a4paper,groupedaddress,preprintnumbers,nofootinbib,floatfix]{revtex4-1}

\usepackage[top=2.9cm, bottom=2.1cm, left=2cm, right=2cm]{geometry}
\usepackage{amsmath,amssymb,amsfonts, dsfont}
\usepackage{graphicx}
\usepackage{color}
\usepackage{hyperref}
\usepackage{subfigure}
\usepackage{dcolumn}% Align table columns on decimal point
\usepackage{bm}% bold math
\usepackage{float}
\usepackage{mathtools}
\usepackage{amsthm}
\usepackage[vcentermath]{youngtab}
\usepackage{cancel} % to make the barred text notation
\usepackage{slashed}
\usepackage{feynmp}
\usepackage{ytableau}
\usepackage{youngtab}
\usepackage{multirow}
\usepackage{feynmp-auto} % For Feynman diagrams. The files are compiled using "mpost <file name>.mp"

\definecolor{orange}{cmyk}{0,0.5,1,0}
\definecolor{rossoCP3}{cmyk}{0,.88,.77,.40}
\definecolor{graa}{rgb}{0.8,0.8,0.8}
\definecolor{blaa}{rgb}{0.2,0.2,0.6}
\hypersetup{
	colorlinks, 
	bookmarksopen, 
	bookmarksnumbered,
	citecolor=blaa, 		%color of links to bibliography
	linkcolor=rossoCP3,	%color of internal links
	urlcolor=rossoCP3,			%color of external links 
	    }
\newcommand{\beq}{\begin{eqnarray}}
\newcommand{\eeq}{\end{eqnarray}}

\newcommand{\SU}{\mathrm{SU}}
\newcommand{\Sp}{\mathrm{Sp}}
\newcommand{\SO}{\mathrm{SO}}
\newcommand{\U}{\mathrm{U}}

\newcommand{\Tr}{\mbox{Tr}}

\begin{document}

%%%%%%%%%%%%%%TITLE AFFILIATIONS ETC%%%%%%%%%%%%%%%%%%%%%%%%%%%%%%%%%%%%%%%%%%%%%%%%%%%%%%%%%%%%
 
\title{\texorpdfstring{\Large  Vacuum alignment in a composite 2HDM }{}}
\author{Chengfeng {\sc Cai}}
%\email{ccfspesysu@gmail.com}
\author{Hong-Hao {\sc Zhang}}
\email{zhh98@mail.sysu.edu.cn}
\affiliation{School of Physics, Sun Yat-Sen University, Guangzhou 510275, China}
\author{Giacomo {\sc Cacciapaglia}}
\email{g.cacciapaglia@ipnl.in2p3.fr} 
\affiliation{Universit\' e de Lyon, F-69622 Lyon, France: Universit\' e Lyon 1, Villeurbanne
CNRS/IN2P3, UMR5822, Institut de Physique Nucl\' eaire de Lyon.}
%%%%%%%%%%%%%%%%%%%%%%%%%%%%%%%%%%%%%%%%%%%%%%%%%%%%%%%%%%%%%%%%%%%%%%%%%%

\begin{abstract}
\vspace{1cm}
We study in detail the vacuum structure of a composite two Higgs doublet model based on a minimal underlying theory with 3 Dirac fermions in pseudo-real representations of the condensing gauge interactions, leading to the $\SU(6)/\Sp(6)$ symmetry breaking pattern. We find that, independently on the source of top  mass, the most general CP-conserving vacuum is characterised by three non-vanishing angles.  A special case occurs if the Yukawas are aligned, leading to a single angle. In the latter case, a Dark Matter candidate arises, protected by a global $\U(1)$ symmetry. 
\\[.3cm]
{\footnotesize  \it Preprint: LYCEN 2017-05}
\end{abstract}
\maketitle
\newpage

%%%%%%%%%%%%%%%%%%%%%%%%%%%%%%%%%%%
\section{Introduction}

Since the discovery of a Higgs boson at the LHC in 2012, the main priority in particle physics has been to establish if the new state really is the last missing particle~\cite{Higgs:1964pj} predicted by the Standard Model (SM) of particle physics. Measuring the properties of the new boson can be seen as a new tool to test theories Beyond the SM.
One very attractive and time-honoured possibility is that the Higgs boson arises as a composite state of a more fundamental confining theory.
This is the only experimentally well tested mechanism of spontaneous symmetry breaking, as it appears at low energies in quantum chromodynamics (QCD).
In fact, as soon as the SM was proposed, models of dynamical symmetry breaking were born~\cite{Weinberg:1975gm,Dimopoulos:1979es}.
While the initial proposals were essentially Higgs-less, i.e. featuring heavy scalar states like in QCD, theories featuring light composite scalars are rather common.
A Higgs-like boson can arise either as an accidentally light scalar resonance (being light, for instance, because of a conformal dynamics at high energies~\cite{Sannino:2004qp,Catterall:2007yx}~\footnote{See also the recent results in Ref.~\cite{Rantaharju:2017eej}.} and/or radiative corrections~\cite{Foadi:2012bb}) or as a pseudo-Nambu-Goldstone boson (pNGB)~\cite{Kaplan:1983fs,Kaplan:1983sm}. 

In this work we are mainly interested in the latter possibility, which received renewed attention in the last decade following the conjectured correspondence between anti-de-Sitter extra dimensions and conformal field theories~\cite{Maldacena:1997re}. Boldly using the conjecture without supersymmetry, the dynamical properties of composite near-conformal theories have been associated with extra dimensional constructions where the Higgs boson arises as an additional polarisation of a gauge boson~\cite{Contino:2003ve}. The phenomenology of composite pNGB Higgses has been widely explored in the recent literature (see for instance the  reviews~\cite{Contino:2010rs,Bellazzini:2014yua,Panico:2015jxa}). The properties of the composite sector mainly depend on the symmetry breaking pattern associated to the strong dynamics. 
A useful guiding principle is thus provided by the definition of a simple underlying gauge-fermion theory that confines in the infra-red, like QCD. 
The existence of a Fundamental Composite Dynamics (FCD)~\cite{Cacciapaglia:2014uja} restricts the cosets to 3 basic possibilities: $\SU(2N)/\Sp(2N)$ for fermions in a pseudo-real representation of the confining gauge group, $\SU(N)/\SO(N)$ for real and $\SU(N)^2/\SU(N)$ for complex ones. It follows that the minimal model is based on $\SU(4)/\Sp(4)$~\footnote{A more minimal coset would be $\SO(5)/\SO(4)$~\cite{Agashe:2004rs}, which however does not have a simple FCD~\cite{Caracciolo:2012je,vonGersdorff:2015fta}. The $\SU(4)/\Sp(4)\sim\SO(6)/\SO(5)$ coset has also been studied in Refs~\cite{Katz:2005au,Gripaios:2009pe,Frigerio:2012uc}.}, which enjoys a very simple FCD based on $G_{\rm TC} = \Sp(2)$ with 2 Dirac fermions in the fundamental representation~\cite{Ryttov:2008xe,Galloway:2010bp} (in the following, for historical reasons, we will denote the underlying dynamics as Technicolor~\cite{Dimopoulos:1979es} -- TC). This theory has also been studied on the lattice, where it has been proven to condense~\cite{Hietanen:2014xca} and its spectrum has been obtained~\cite{Drach:2015epq,Arthur:2016dir,Arthur:2016ozw}. An analysis based on a Nambu-Jona-Lasinio modelling can be found in Ref.~\cite{Bizot:2016zyu}.
Also, minimality can now be defined not in terms of the number of pNGBs the coset contains (aka ``coset-ology''), but in terms of the degrees of freedom of the underlying theory.

Going non-minimal has its own advantages: having more light pNGBs allows to study more complex Higgs sectors. For instance, a composite two Higgs doublet model (2HDM) can be obtained from a QCD-like $G_{\rm TC} =\SU(N)$ theory with 4 fundamental Dirac fermions~\cite{Ma:2015gra} (for other composite 2HDMs, see Refs~\cite{Mrazek:2011iu,Bertuzzo:2012ya,DeCurtis:2016gly}). The additional pNGBs can also play the role of Dark Matter~\cite{Frigerio:2012uc} if the coset allows for a protecting symmetry~\cite{Chala:2012af,Ma:2017vzm,Ballesteros:2017xeg,Balkin:2017aep}.
In this work we are interested in an extension of the minimal FCD model based on the coset $\SU(6)/\Sp(6)$, which can be minimally obtained in an $\Sp(2)$ model with 3 Dirac fermions and leads naturally to a composite 2HDM. The main difference with the FCD in Ref.~\cite{Ma:2015gra}, based on $\SU(4)^2/\SU(4)$, is about the action of the custodial symmetry on the two doublets. Including a custodial symmetry in the global symmetry is crucial in order to obtain realistic models~\cite{Georgi:1984af}. In models with two doublets, the custodial symmetry can be defined in two inequivalent ways~\cite{Pomarol:1993mu,Grzadkowski:2010dj}: in the model of Ref.~\cite{Ma:2015gra} the two doublets transform as a complex bi-doublet under the gauged $\SU(2)_L$ and the global $\SU(2)_R$. We will see that in the $\SU(6)/\Sp(6)$ model under consideration one can define two global $\SU(2)_R$'s acting, individually, on the two doublets.

We finally remark that an underlying theory similar to the one we are interested in has been used in Refs~\cite{Low:2002ws,Brown:2010ke} to construct a Little Higgs model, where the light Higgs also arises as a pNGB from the $\SU(6)/\Sp(6)$ coset. The main difference between the two approaches is that in Little Higgs models the gauged subgroup is more extended than the electroweak (EW) one, and the additional gauge generators are assumed to be broken at the condensation scale.

%%%%%%%%%%%%%%%%%%%%%%%%%%%%%%%%%%%%
\subsection{Preliminaries: $\SU(6) \to \Sp(6)$}

The unbroken group $\Sp(6)$, of rank 3, contains 3 commuting $\SU(2)$ subgroups: it is thus natural to try to identify them with the EW sector of the SM, i.e. the gauged $\SU(2)_L$ and the partly-gauged $\SU(2)_R$. We insist on being able to identify an $\SU(2)_R$ in order to obtain a model that respects a custodial symmetry.
Furthermore, the 14 Goldstone bosons in the coset transform, under the 3 $\SU(2)$'s, as:
\beq
{\bf 14}_{\Sp(6)} \to (2,2,1) \oplus (2,1,2) \oplus (1,2,2) \oplus (1,1,1) \oplus (1,1,1)\,.
\eeq
A composite Higgs boson can, thus, be identified with any one of the 3 bi-doublets. To classify the various possibilities, it is useful to list the quantum number assignments of a fundamental $\bf 6$ of $\SU(6)$, $\psi$, that will correspond to the techni-fermions in the underlying gauge-fermion model.
Each sextet $\psi$ can be decomposed into 3 doublets $\psi_i$, $i=1,2,3$, each one transforming under a different global $\SU(2)_i$.
Note that $\psi$ is a left-handed Weyl spinor, and each doublet $\psi_i$ forms a Dirac spinor (which can be assigned an $\SU(2)$-invariant mass).
There are, therefore, 2 possible assignments that lead to a custodial invariant composite Higgs, whose properties are listed in Table~\ref{tab:models}. 
\begin{table}[tb!]
\begin{tabular}
{c|c|cc|c|c|c|}
 & & $\SU(2)_L$ & $\U(1)_Y$ & $\SU(2)_L$ & $Y$ & Higgs\\
 \hline
\multirow{3}{*}{Case A \phantom{xxx}} & $\psi_1$ & \textbf{2} & $0$ & \multirow{3}{*}{SU(2)$_1$} & \multirow{3}{*}{$T^3_{2} + \xi T^3_{3}$} & \multirow{3}{*}{$\begin{array}{c} (2,2,1) \\ \left[(2,1,2)\; \mbox{if}\; \xi=1\right]\end{array}$}\\
& $\psi_2$ & \textbf{1} & $\pm 1/2$ &&&\\
& $\psi_3$  & \textbf{1} & $\pm \xi/2$ &&&\\
 \hline
\multirow{3}{*}{Case B\phantom{xxx}} & $\psi_1$ & \textbf{2} & $0$ & \multirow{3}{*}{SU(2)$_1$ + SU(2)$_2$} & \multirow{3}{*}{$T^3_{3}$} & \multirow{3}{*}{$(2,1,2) + (1,2,2)$}\\
& $\psi_2$ & \textbf{2} & $0$ &&&\\
& $\psi_3$  & \textbf{1} & $\pm 1/2$ &&&\\
 \hline
 \end{tabular}
 \caption{Quantum number assignments for the underlying techni-fermions. $T^3_i$ are the diagonal generators of each $\SU(2)_i$.} \label{tab:models}
 \end{table}

In Case A, the pNGBs in the multiplets $(1,2,2)$ and $(2,1,2)$ contain scalars with electromagnetic charges $Q = \pm \frac{1\pm \xi}{2}$, so that only choosing integer odd values for $\xi$ avoids states with non-integer charges. In particular, for $\xi=1$, the model contains two Higgs doublets. 

In Case B, there are always 2 Higgs doublets, however the $(2,2,1)$ pNGBs contain an isotriplet: if the vacuum misalignment projects a vacuum expectation value on the triplet, the custodial symmetry would be broken, thus incurring in strong constraints. Furthermore, the contribution to precision tests, in particular to the $S$ parameter, is sensitive to the number of EW doublets~\cite{Peskin:1991sw}, thus one would expect milder constraints in Case A where the underlying theories contain only one doublet.

For the above reasons, in the following we will focus on the family of models in Case A, discussing in particular the case $\xi=1$ that provides a composite 2HDM.
The paper is organised as follows: in Section~\ref{sec:model} we introduce the model and study the simplest vacuum alignment. We comment on various mechanisms that generate the top mass, and find that the only consistent configuration entails misalignment along one of the two Higgs doublets and a Dark Matter candidate protected by a global $\U(1)$ symmetry. Then, in Section~\ref{sec:vacuum}, we study a more general vacuum alignment, naturally twisted along a singlet direction. We present some numerical results, mainly focussed on the mass of the Higgs boson candidate, before presenting our conclusions and perspectives in Section~\ref{sec:concl}.

%%%%%%%%%%%%%%%%%%%%%%%%%%%%%%%%%%%%%%%%%%
\section{The Model} \label{sec:model}

The underlying model consists of 3 Dirac fermions $\psi_i$ in a pseudo-real representation of the strongly interacting group $G_\text{TC}$, that we expect to condense and confine in the infrared.
The minimal model, in terms of degrees of freedom, respecting these properties is $G_{\rm TC} = \SU(2)$ with fermions in the fundamental.
The global symmetry of the strong sector is $\SU(6)$ (times an anomalous $\U(1)$): the EW gauge symmetry is embedded as in Case A with $\xi=1$, i.e. $\SU(2)_L$ is identified with $\SU(2)_1$ acting on $\psi_1$, while the hypercharge is identified with the diagonal generator of the two remaining $\SU(2)$'s.
The pions (pNGBs) generated by the symmetry breaking $\SU(6) \to \Sp(6)$ include two Higgs doublet candidates transforming under $\SU(2)_{R1} \equiv \SU(2)_2$ and $\SU(2)_{R2} \equiv \SU(2)_3$ respectively.
One can thus identify two custodial symmetries, depending on which bi-doublet breaks the EW symmetry:
\beq
\underbrace{\SU(2)_L \times \SU(2)_{R1}}_{\SO(4)_1} \times \SU(2)_{R2} \qquad \text{or} \qquad \underbrace{\SU(2)_L \times \SU(2)_{R2}}_{\SO(4)_2} \times \SU(2)_{R1}\,.
\eeq
In general, the vacuum can be misaligned along both doublets, thus care is needed in order to preserve custodial symmetry.

The underlying Lagrangian for the techni-fermions thus reads:
\begin{equation}
\mathcal{L} = i \bar\psi \sigma^\mu D_\mu \psi - \psi \mathcal{M}_\psi \psi - \bar{\psi} \mathcal{M}_\psi^\dagger \bar{\psi}\,,
\end{equation}
where $\psi$ is a 6-component Weyl spinor, and $\mathcal{M_\psi}$ is a mass term that can be written, without loss of generality, as:
\begin{equation} \label{eq:MQ}
\mathcal{M_\psi} = \begin{pmatrix}
m_L (i\sigma_2) & 0 & 0 \\
0 & m_{R1} (-i \sigma_2) & 0\\
0 & 0 & m_{R2} (-i \sigma_2)
\end{pmatrix}\,.
\end{equation}
The signs of the mass terms have been chosen arbitrarily, following the choice for the vacuum alignment explained below, and we also chose to work in the basis where the masses of the two $\SU(2)_L$ singlets are diagonal. This mass term breaks $\SU(6) \to \Sp(2)^3$, however the symmetry is enhanced to $\Sp(2)\times\Sp(4)$ if two masses are equal and to $\Sp(6)$ is all masses are equal (recall that $\Sp(2)\sim\SU(2)$).

The covariant derivative contains both the FCD gauge interactions and the EW ones. The latter ones are embedded in the following way : 
\begin{equation}
      T^i_L = \underbrace{\frac{1}{2}
      \begin{pmatrix}
      \sigma_i & 0 & 0 \\
      0 & 0 & 0 \\
      0 & 0 & 0
      \end{pmatrix}}_{\SU(2)_L \text{ generators}} \ , \ \quad
      T^i_{R1} = \underbrace{\frac{1}{2}
      \begin{pmatrix}
      0 & 0 & 0 \\
      0 & -\sigma_i^T & 0 \\
      0 & 0 & 0
      \end{pmatrix}}_{\SU(2)_{R1} \text{ generators}} \quad \text{ and }  \quad
      T^i_{R2} = \underbrace{\frac{1}{2}
      \begin{pmatrix}
      0 & 0 & 0 \\
      0 & 0 & 0 \\
      0 & 0 & -\sigma_i^T
      \end{pmatrix}}_{\SU(2)_{R2}\text{ generators}}\ ,
       \qquad i=1,2,3\,,
\end{equation}
where $Y = T_{R1}^3 + T_{R2}^3$ is identified with the hypercharge generator.
The above choice (under which $\psi_2$ and $\psi_3$ are actually anti-doublets of the corresponding $\SU(2)$'s) is consistent with the following choice for the condensate alignment that preserves the EW symmetry:
\begin{equation}
\left< \psi \psi \right> \sim \Sigma_0 = \begin{pmatrix}
i\sigma_2 & 0 & 0\\
0 & -i\sigma_2 & 0 \\
0 & 0 & - i \sigma_2 
\end{pmatrix}\,.
\end{equation}
A complete list of the 14 broken generators $X^i$ and of the 21 unbroken ones $S^j$ can be found in Appendix~\ref{app:generators}.  
Under the EW symmetry $\SU(2)_{L} \times \U(1)_Y$, the pNGBs transform as
\begin{equation}
\mathbf{14}_{\Sp(6)} = \mathbf{2}_{\pm 1/2}+\mathbf{2}_{\pm 1/2}+\mathbf{1}_{\pm 1} + \mathbf{1}_{0} + \mathbf{1}_{0} + \mathbf{1}_{0} + \mathbf{1}_{0}\,,
\end{equation}
where the charged and 2 neutral singlets belong to the bi-doublet under the $\SU(2)_R$ symmetries.
The model contains three bi-doublets but only two are doublets of $\SU(2)_L$ and may play the role of a Brout-Englert-Higgs doublet.
The pion matrix is embedded in
\begin{equation} \label{eq:U0}
U_0 (\phi) = e^{i \Pi(\phi)/f}\,, \qquad \mbox{with} \quad \Pi(\phi) = \sum_{i=1}^{14} \phi_i X^i\,.
\end{equation}
One can now perturb the EW preserving vacuum $\Sigma_0$ along the broken directions to make the pNGBs appear
\begin{equation}
\Sigma_0(\phi) = U_0 (\phi)\cdot \Sigma_0\,.
\end{equation}
The object we defined above, which transforms linearly under $\SU(6)$, can be used to define the chiral Lagrangian 
\begin{equation} \label{eq:Lchi2}
\mathcal{L}_{(p^2)} = f^2 \Tr \left[ (D_\mu \Sigma (\phi))^\dagger \cdot D^\mu \Sigma (\phi) \right] - \Tr \left[ \chi \cdot \Sigma^\dagger (\phi) + \chi^\dagger \cdot \Sigma (\phi) \right]\,,
\end{equation}
where the spurion $\chi$ contains the techni-fermion mass term:
\begin{equation}
\chi = 2 B M_\psi^\dagger\,,
\end{equation}
$B$ being a low energy constant that can be computed on the lattice.

%%%%%%%%%%%%%%%%%%%%%%%%%%
\subsection{Vacuum misalignment}

The EW symmetry breaking is triggered, in this model, by the pNGBs acquiring a vacuum expectation value (VEV). In general, all the scalars can develop it, however only two correspond to the real part of the neutral components of the two Higgs doublets:
\begin{equation}
\langle \phi_4 \rangle = v_1\,, \quad \langle \phi_8 \rangle = v_2\,.
\end{equation}
The potential for the pNGBs is generated by all explicit $\SU(6)$ breaking terms. Our strategy will be to apply the ansatz that only the two above scalars are involved, and then check under which conditions this is realised by the potential.

A pion acquiring a VEV generates a misalignment of the vacuum, i.e. it is possible to define an $\SU(6)$ transformation that rotates the vacuum $\Sigma_0$ to the true minimum of the theory. 
In our case, this is defined  by a rotation $\Omega (\theta, \beta)$, such that
\begin{equation}
\Sigma_{\theta, \beta} = \Omega (\theta, \beta) \cdot \Sigma_0 \cdot \Omega^T (\theta, \beta)\,,
\end{equation}
where $\tan \beta = v_2/v_1$ and $\theta = \frac{\sqrt{v_1^2+v_2^2}}{2 \sqrt{2} f}$.
Note that the pion matrix on the new vacuum is defined as
\begin{equation}
\Pi (\phi)_{\theta, \beta} = \Omega \cdot \Pi (\phi) \cdot \Omega^\dagger\,, \quad U (\phi)_{\theta, \beta} = \Omega \cdot U_0 (\phi) \cdot \Omega^\dagger\,,
\end{equation}
and
\begin{equation}
\Sigma (\phi) = U_{\theta, \beta} (\phi)\cdot \Sigma_{\theta, \beta} = \Omega\cdot U_0 (\phi)\cdot \Sigma_0\cdot \Omega^T\,.
\end{equation}

The rotation $\Omega$ that applies to our ansatz can be decomposed as:
\begin{equation} \label{eq:Omegatb}
\Omega (\theta, \beta) =  R_\beta \cdot \Omega_\theta \cdot R_\beta^\dagger\,,
\end{equation}
where $\Omega_\theta$ is a rotation generated by $X^4$:
\begin{equation} \label{eq:Omega}
\Omega_\theta = e^{i \sqrt{2} \theta\ X^4} = \begin{pmatrix}
\cos{\frac{\theta}{2}} \mathbb{I}_2 & \sin{\frac{\theta}{2}} \; i\sigma_2 & 0\\
\sin{\frac{\theta}{2}}  \; i\sigma_2 & \cos{\frac{\theta}{2}}\ \mathbb{I}_2 & 0  \\
0 & 0 & \mathbb{I}_2
\end{pmatrix}\,.
\end{equation}
The rotation $R_\beta$ is generated by $S^{21}$: 
\begin{equation} \label{eq:Omegabeta}
R_\beta = e^{i 2\sqrt{2}\beta \ S^{21}} = \begin{pmatrix}
\mathbb{I}_2 & 0 & 0\\
0 & \cos{\beta}\ \mathbb{I}_2 & -\sin{\beta}\ \mathbb{I}_2 \\
0 & \sin{\beta}\ \mathbb{I}_2 & \cos{\beta}\ \mathbb{I}_2
\end{pmatrix}\,.
\end{equation}
The fact that the rotation $R_\beta$ is generated by an unbroken generator of the vacuum $\Sigma_0$, allows to write the pNGB-dependent vacuum matrix as
\begin{equation} \label{eq:RbetaRot}
\Sigma (\phi) = R_\beta \cdot \Omega_\theta \cdot U_0 (\phi') \cdot \Sigma_0\cdot \Omega_\theta^T \cdot R_\beta^T\,,
\end{equation}
where $U_0 (\phi') = R_\beta^\dagger \cdot U_0 (\phi)\cdot R_\beta$. The rotation $R_\beta$  reshuffles the broken generators, so $U_0 (\phi')$ is simply a change of basis, furthermore it drops out from the first term of the chiral Lagrangian in Eq.(\ref{eq:Lchi2}) as the gauged generators are left invariant. 
It is therefore convenient to study the theory in a vacuum that only depends on $\theta$, with the pNGB basis $\phi'$, so that in the following
\begin{equation} \label{eq:Sigmafinal}
\Sigma (\phi) = \Omega_\theta \cdot U_0 (\phi') \cdot \Sigma_0\cdot \Omega_\theta^T\,.
\end{equation}
 All the spurions that explicitly break $\SU(6)$ thus need to be rotated by $R_\beta^\dagger$: if they are invariant, the $\beta$-dependence will be completely removed, indicating that the vacuum of the theory can be characterised in terms of a single angle; if not, the dependence on $\beta$ will be carried-off by the spurions themselves.

The pNGB matrix can be conveniently expressed as follows:
\begin{equation} \label{eq:pNGBmatrix}
i \Pi(\phi')\cdot \Sigma_0  = \frac{1}{2} \left(
\begin{array}{ccc}
- \left(\frac{1}{\sqrt{2}} \eta_1 + \frac{1}{\sqrt{6}} \eta_2 \right) \sigma^2 & H_1 & H_2 \\
- H_1^T & -\left( \frac{1}{\sqrt{2}} \eta_1 - \frac{1}{\sqrt{6}}  \eta_2 \right) \sigma^2 & G \\
- H_2^T & - G^T & - \sqrt{\frac{2}{3}} \eta_2 \sigma^2
\end{array}
\right)
\end{equation}
with $\eta_{1,2}$ being the singlet pseudo-scalars $(1,1,1)$, and $H_1$, $H_2$, $G$ being, respectively, the $\SU(2)_L \times \SU(2)_{R1}$, $\SU(2)_L \times \SU(2)_{R2}$ and $\SU(2)_{R1} \times \SU(2)_{R2}$ bi-doublets. In components, they can be written as: 
\begin{eqnarray}
& H_1 = \frac{1}{\sqrt{2}}\begin{pmatrix}
 h_1 + i \pi_3 & i \pi_1 + \pi_2 \\
 i \pi_1 - \pi_2 & h_1 - i \pi_3 \\
\end{pmatrix}\,, 
\qquad H_2 = \frac{1}{\sqrt{2}}\begin{pmatrix}
 h_2 + i A_0 & \sqrt{2} H^+ \\
 - \sqrt{2} H^- & h_2 - i A_0 \\
\end{pmatrix}\,, & \nonumber \\
%\end{equation}
%\begin{equation} 
& G = \frac{1}{\sqrt{2}}\begin{pmatrix}
- \sqrt{2} \eta^- & \varphi_0 - i \eta_3 \\
 \varphi_0 + i \eta_3 & \sqrt{2} \eta^+ \\
\end{pmatrix}\,. &
\end{eqnarray}
In this basis, $H_1$ is the doublet that develops a VEV and breaks the EW symmetry, so that $h_1$ will play the role of the pNGB Higgs and $\pi_i$ are the exact Goldstones eaten by the massive $W$ and $Z$, while $H_2$ is the second doublet. $G$, on the other hand, contain two neutral singlets ($\eta_3$ being a pseudo-scalar) and a charged one.
More details on the CP properties of the pNGBs are provided in Appendix~\ref{app:CP}, while the lowest order pNGB couplings, in the new basis, are given in Appendix~\ref{app:pNGBcoupl} and explicitly show no dependence on $\beta$.
The LO chiral Lagrangian generates the following masses for the $W$ and $Z$:
\begin{equation}
 	  m_W^2 = 2 g^2 f^2 \sin^2\theta \; , \qquad m_Z^2 = \frac{m^2_W}{\cos^2{\theta_W}}\,,
\end{equation}
so that
\begin{equation}
 	  v_{SM} = 2\sqrt{2} f \sin\theta = 246\; \text{GeV}\,
\end{equation}
fixes the relation between the decay constant $f$ and the EW scale via the alignment angle $\theta$.
The field $h_1$ is the only one that has linear couplings to the vector bosons and thus can play the role of the Higgs boson. Its couplings are given by:
\begin{equation}
 	  g_{h_1WW}\ = \ g_{h_1ZZ} \cos^2(\theta_W) \ = \ \sqrt{2} g^2 f \sin\theta \cos\theta \ = \ g_{hWW}^{\rm SM} \cos\theta\,. \label{eq:ghWW}
\end{equation}
The results above match the ones from more minimal models~\cite{Agashe:2004rs,Cacciapaglia:2014uja}.

The second term in Eq.(\ref{eq:Lchi2}), which contains the techni-fermion mass, can also be written in terms of the $\theta$-vacuum following Eq.(\ref{eq:Sigmafinal}) by defining a rotated mass: 
\begin{equation}
\mathcal{M}'_\psi = R_\beta^\dagger \cdot \mathcal{M}_\psi \cdot R_\beta = \begin{pmatrix}
m_L (i \sigma^2) & 0 & 0 \\
0 & (m_R + \delta m_R c_{2\beta}) (-i \sigma^2) & \delta m_R s_{2\beta} (i \sigma^2) \\
0 & \delta m_R s_{2\beta} (i \sigma^2) & (m_R - \delta m_R c_{2\beta}) (-i \sigma^2)
\end{pmatrix}\,,
\end{equation}
where we employ the compact notation $s_x = \sin x$ and $\cos x = c_x$, and we defined 
\begin{equation}
m_R = \frac{m_{R1} + m_{R2}}{2}\,, \quad \delta m_R = \frac{m_{R1} - m_{R2}}{2}\,.
\end{equation}
We see that the $\beta$-dependence drops if $\delta m_R = 0$, i.e. when the two $\SU(2)_R$ doublets are degenerate.

The misalignment of the vacuum, and the masses of the remaining pNGBs, are determined by the effective potential, that is dominated by the effect of loops of tops, i.e. by the explicit breaking of the global symmetry introduced by the interactions giving rise to the mass of the top.
In the following, we will first discuss the effect of the effective top (and bottom) Yukawas, before presenting general results on the potential.

%%%%%%%%%%%%%%%%%%%%%%%%%%%%%%%%%%%%%%
\subsection{Finding a minimum: the top (and bottom) Yukawas} \label{sec:Yuktopbot}

As a warm up, we will first consider the top mass generated by 4-fermion interactions, bilinear in the SM fields $Q_L$ and $t_R^c$ (defined here as left-handed Weyl spinors):
\begin{equation}
\frac{y'_{t1}}{\Lambda_t^2}\ \left( Q_L t_R^c \right)^\dagger_\alpha\  (\psi^T P_1^\alpha  \psi) + \frac{y'_{t2}}{\Lambda_t^2}\ \left( Q_L t_R^c \right)^\dagger_\alpha\ (\psi^T P_2^\alpha  \psi)\,,
\end{equation}
where $ \alpha$ is an $\SU(2)_L$ index and $P_1^\alpha$, $P_2^\alpha$ are projectors (defined in  Appendix~\ref{app:topproj}) that select the components of $\langle \psi^T \psi \rangle$  transforming as the Higgs doublets. This is a generalisation of the four-fermion interactions introduced in Refs~\cite{Galloway:2010bp,Cacciapaglia:2014uja}. The couplings $y'_{t1}$ and $y'_{t2}$ are the pre-Yukawas, while $\Lambda_t$ is a new dynamical scale where the 4-fermion interactions are generated.
This scenario is inspired by the old Extended Technicolor~\cite{Eichten:1979ah} idea, where such interactions were generated by an extended gauge sector. A generic problem with this approach, that renders it somewhat unlikely, is the fact that other four-fermion interactions, including flavour violating ones in the SM fields, are generated at the same scale. This is a problem as $\Lambda_t$ cannot be too large without suppressing the top (and other quarks') mass, unless the $\psi^T \psi$ operators has a large anomalous dimension~\footnote{The tension is in the fact that the bilinear operator $\psi^T \psi$, that enters in the top mass, needs a dimension close to 1 (i.e. that of an elementary scalar field), while the dimension of $(\psi^T \psi)^2$, that induces a mass for the pNGB Higgs, needs to be close to 4. Bootstrap techniques have been used to prove the existence of an upper limit on the anomalous dimensions of scalar operators that seem to disprove the above situation~\cite{Rattazzi:2008pe,Rychkov:2009ij}. However, the bound applies to the anomalous dimension of the operator with smallest dimension, which may not be the one associated to the Higgs mass (see Ref.~\cite{Antipin:2014mga} for a counter-example). The feasibility of this scenario is, therefore, still an open question.}. One easy way out would be to consider models where the top mass is generated at a different scale from the other quarks and leptons: this would be enough to suppress dangerous flavour violation~\cite{Cacciapaglia:2015dsa,Panico:2016ull}, however complete models of this kind are still lacking (for a bookkeeping example, see~\cite{Cacciapaglia:2015yra}). In a following subsection we will discuss how other mechanisms generating fermion masses would affect the conclusions we reach in this section. For now, we take this simple mechanism for the top, without worrying about the full flavour structure, and analyse the impact on the vacuum alignment.

First, we note that the effect of the two pre-Yukawas can be embedded into a single spurion that is a linear combination of the two projectors, i.e. $y_{t1} P_1^\alpha + y_{t2} P_2^\alpha$, where we removed the prime to distinguish the couplings of the effective chiral Lagrangian from the couplings of the four-fermion interactions, to which they are related via form factors. To define the theory in the $\theta$-vacuum of Eq.(\ref{eq:Sigmafinal}), we need to rotate the spurion as follows:
\begin{equation}
R_\beta \cdot (y_{t1} P_1^\alpha + y_{t2} P_2^\alpha) \cdot R_\beta^\dagger = Y_{t1} P_1^\alpha + Y_{t2} P_2^\alpha\,,
\end{equation}
where
\begin{equation} \label{eq:Yt1Yt2}
Y_{t1}  =  c_\beta y_{t1} + s_\beta y_{t2} \,, \quad  Y_{t2}  =  -s_\beta y_{t1} + c_\beta y_{t2}\,,
\end{equation}
carry the dependence on $\beta$ (we remark that $\frac{\partial}{\partial \beta} Y_{t1} = Y_{t2}$ and $\frac{\partial}{\partial \beta} Y_{t2} = - Y_{t1}$).
The operator in the chiral Lagrangian generating the top mass is thus:
\begin{multline} \label{eq:topmass}
\mathcal{L}_{(p^2)} \supset f\ \left( Q_L t_R^c \right)^\dagger_\alpha\ \left( Y_{t1} \text{Tr} [P^\alpha_1\cdot \Sigma (\phi') ] + Y_{t2} \text{Tr} [P^\alpha_2\cdot \Sigma (\phi')] \right) \sim 
- f \sin \theta\ Y_{t1}\ \left( t_L t_R^c \right)^\dagger \\
- \left[\frac{Y_{t1}}{2\sqrt{2}} \left( c_\theta\ h_1 + \frac{i}{\sqrt{3}} s_\theta\ \eta_2 \right) + \frac{Y_{t2}}{2\sqrt{2}} \left(c_\frac{\theta}{2}\ (h_2 + i A_0) + s_{\frac{\theta}{2}} (\varphi_0 + i \eta_3)  \right)  \right] \left( t_L t_R^c \right)^\dagger \\
+ \left[ \frac{Y_{t1}}{2} c_{\frac{\theta}{2}}\ H^- - \frac{Y_{t2}}{2} s_{\frac{\theta}{2}}\ \eta^-  \right]  \left( b_L t_R^c \right)^\dagger + \dots
\end{multline}
where the dots include couplings to the pNGBs. 
The top mass is thus proportional to $Y_{t1}$ ($m_t = Y_{t1} f \sin \theta$), while $Y_{t2}$ generates couplings of other pNGBs.
The mass for the bottom quark can be generated by a very similar operator, defined in terms of two other projectors $P_{1b}$ and $P_{2b}$ (also defined in  Appendix~\ref{app:topproj}): 
\begin{multline} \label{eq:botmass}
\mathcal{L}_{(p^2)} \supset f\ \left( Q_L b_R^c \right)^\dagger_\alpha\ \left( Y_{b1} \text{Tr} [P^\alpha_{1b}\cdot \Sigma (\phi') ] + Y_{b2} \text{Tr} [P^\alpha_{2b}\cdot \Sigma (\phi')] \right) \sim 
- f \sin \theta\ Y_{b1}\ \left( b_L b_R^c \right)^\dagger \\
- \left[\frac{Y_{b1}}{2\sqrt{2}} \left( c_\theta\ h_1 + \frac{i}{\sqrt{3}} s_\theta\ \eta_2 \right) + \frac{Y_{b2}}{2\sqrt{2}} \left(c_\frac{\theta}{2}\ (h_2 - i A_0) - s_{\frac{\theta}{2}} (\varphi_0 - i \eta_3)  \right)  \right] \left( b_L b_R^c \right)^\dagger \\
- \left[ \frac{Y_{b1}}{2} c_{\frac{\theta}{2}}\ H^+ + \frac{Y_{b2}}{2} s_{\frac{\theta}{2}}\ \eta^+  \right]  \left( t_L b_R^c \right)^\dagger + \dots
\end{multline}
where the $\beta$-dependence is encoded in
\begin{equation} \label{eq:Ydef1}
Y_{b1}  =  c_\beta y_{b1} + s_\beta y_{b2} \,, \quad  Y_{b2}  =  -s_\beta y_{b1} + c_\beta y_{b2}\,.
\end{equation}
Couplings of two pNGBs to the fermions, generated by the Yukawa operators, are reported in Appendix~\ref{app:pNGBcoupl2}.

The contribution of the Yukawas to the potential for the vacuum alignment is thus given by:
\begin{equation}
V_{\rm Yuk} = - C_t f^4 \sum_\alpha \left( \left| Y_{t1} \text{Tr} [P^\alpha_1\cdot \Sigma  ] + Y_{t2} \text{Tr} [P^\alpha_2\cdot \Sigma]\right|^2 + \left| Y_{b1} \text{Tr} [P^\alpha_{b1}\cdot \Sigma ] + Y_{b2} \text{Tr} [P^\alpha_{b2}\cdot \Sigma ]  \right|^2  \right)\,.
\end{equation}
Expanding the above operator up to linear terms in the pNGBs: 
\begin{multline} \label{eq:Vtop}
V_{\rm Yuk} = - C_t f^4 \left\{ \left( |Y_{t1}|^2 + |Y_{b1}|^2 \right) s^2_\theta + \frac{h_1}{2 \sqrt{2} f}  \left( |Y_{t1}|^2 + |Y_{b1}|^2 \right) s_{2 \theta} + \right. \\
\left. \frac{h_2}{\sqrt{2} f} \left( \Re\ Y_{t1} Y_{t2}^\ast + \Re\ Y_{b1} Y_{b2}^\ast \right) c_\frac{\theta}{2} s_\theta + \frac{\varphi_0}{\sqrt{2} f} \left( \Re\ Y_{t1} Y_{t2}^\ast - \Re\ Y_{b1} Y_{b2}^\ast \right) s_\frac{\theta}{2} s_\theta + \right. \\
\left. \frac{A_0}{\sqrt{2} f} \left( \Im\ Y_{t1} Y_{t2}^\ast - \Im\ Y_{b1} Y_{b2}^\ast \right) c_\frac{\theta}{2} s_\theta + \frac{\eta_3}{\sqrt{2} f} \left( \Im\ Y_{t1} Y_{t2}^\ast + \Im\ Y_{b1} Y_{b2}^\ast \right) s_\frac{\theta}{2} s_\theta\right\}\,,
\end{multline}
where $\Re$ and $\Im$ indicate, respectively, the real and imaginary parts.
The above form of the potential can be generalised to including all SM fermions by summing over the respective Yukawa couplings. \footnote{It is enough to replace $Y_{ti} Y_{tj}^\ast \to \sum_u Y_{ui} Y_{uj}^\ast$, where the sum included the 3 up-type quarks, and $Y_{bi} Y_{bj}^\ast \to \sum_d Y_{di} Y_{dj}^\ast + \frac{1}{3} \sum_l Y_{li} Y_{lj}^\ast $, with sums over the down-type quarks and leptons (the factor of $3$ compensating the colour multiplicity).}
The potential generates tadpoles for many pNGBs, as a sign that the vacuum alignment is not well defined. Some of them are, however, easily accounted for: the tadpole for the Higgs-like state $h_1$ vanishes once the potential is minimised in $\theta$, while the one for $h_2$ vanishes after minimisation w.r.t. $\beta$, as it is proportional to $\frac{\partial}{\partial \beta} V_{\rm Yuk}^{(0)}$. Hence, the cancellation of the $h_2$ tadpole determines $\beta$ in terms of the pre-Yukawa couplings.
On the other hand, the tadpoles for $A_0$, $\eta_3$ and $\varphi_0$ do not automatically vanish: either the Yukawas are aligned in such a way that the coefficient vanish at the minimum for $\theta$ and $\beta$, or the vacuum needs to be misaligned along the generators associated to the 3 pNGBs.

Both tadpoles for the pseudo-scalars $A_0$ and $\eta_3$ are proportional to residual phases in the pre-Yukawas, thus it would be enough to assume that the two couplings have the same phases for them to vanish. Note that misaligning the vacuum along $A_0$ is particularly dangerous, as it leads to a breaking of the custodial symmetry (and a contribution to the $\rho$ parameter, like in standard 2HDMs~\cite{Pomarol:1993mu,Grzadkowski:2010dj}).
The vanishing of the tadpole for the scalar $\varphi_0$, on the other hand, would require a non-trivial relation between the real parts of the Yukawa couplings:
\begin{equation} \label{eq:alignment}
\Re\ Y_{t1} Y_{t2}^\ast = \Re\ Y_{b1} Y_{b2}^\ast = 0\,, \qquad \Rightarrow \quad \tan \beta = \frac{y_{t2}}{y_{t1}} = \frac{y_{b2}}{y_{b1}}\,,
\end{equation}
where the latter relation is valid if the pre-Yukawas have the same phase (i.e., if the tadpoles for $A_0$ and $\eta_3$ also vanish). In other words, a sufficient condition for the vanishing of the additional tadpoles is that the pre-Yukawas are aligned, i.e. each SM fermion couples to the same combination of the two $\SU(2)_R$ fermions.
Then, the rotation $R_\beta$ would tell us that the doublet that develops the EW-breaking VEV is the same one that couples to all SM fermions.

For completeness, we also add the contribution to the misalignment from gauge interactions, and from the mass term of the techni-fermions in Eq.~(\ref{eq:MQ}).
The latter yields:
\begin{multline} \label{eq:Vm}
V_{\rm m} = 2 B\ \text{Tr} [\mathcal{M}'_\psi\cdot  \Sigma] + h.c. = - 8 B \left\{ M\ (1-\Delta + 2  c_\theta) - \delta m_R\ c_{2\beta}\ (1- c_\theta) \phantom{\frac{h_2}{\sqrt{2} f}} \right.\\
\left. - \frac{h_1}{2 \sqrt{2} f} \left(2 M +\delta m_R c_{2\beta} \right) s_\theta + \frac{h_2}{\sqrt{2} f} \delta m_R s_{2\beta} s_\frac{\theta}{2} + \dots\right\}
\end{multline}
where we have expanded up to linear terms in the pNGBs, and defined
\begin{equation} \label{eq:MDelta}
M = \frac{m_L + m_{R}}{2}\,, \quad \Delta = \frac{m_L - m_R}{m_L + m_R}\,.
\end{equation}
Besides the tadpole for the Higgs related to the minimisation of the potential for $\theta$, a tadpole for $h_2$ emerges proportional to the mass difference between the two Dirac $\SU(2)_L$ singlets.
In presence of $\delta m_R \neq 0$, therefore, the value of $\beta$ at the minimum is modified by the presence of the potential above and it is not possible any more to cancel the tadpole for $\varphi_0$ with a simple alignment of the top and bottom pre-Yukawas, like in Eq.~(\ref{eq:alignment}).

Finally, one needs to consider the contribution of the gauge loops:
\begin{multline} \label{eq:Vgauge}
V_g  = -C_g f^4 \left\{ g^2\ \text{Tr}\left[ T^i_L \Sigma \left( T^i_L \Sigma \right)^* \right] + g'^2  \text{Tr}\left[ Y \Sigma \left( Y \Sigma \right)^* \right]  \right\} = \\
 = - C_g f^4 \left\{ \frac{g'^2}{2} +  \frac{3 g^2 + g'^2}{2} c^2_\theta 
- \frac{h_1}{2 \sqrt{2}} \frac{ 3g^2+g'^2}{2}  s_{2\theta}  + \dots \right\}\,.
\end{multline}
This term will only affect the minimisation for $\theta$. Furthermore, it does not depend on $\beta$ nor it will change our conclusions on the tadpole cancellations for other pNGBs.

\subsubsection*{Summary of findings}

In summary, we have found that it is not enough to describe the misalignment in terms of the EWSB angle $\theta$ and of $\beta$. While choosing real pre-Yukawas (more precisely, it suffices that $y_{f1} y_{f2}^\ast$ is real for all SM fermions so that the SM phase of the CKM matrix can be accommodated) allows us to avoid tadpoles for the pseudo-scalars $A_0$ and $\eta_3$, it is not possible, in general, to eliminate the tadpole for the scalar $\varphi_0$. In the next section, we will therefore define a more appropriate vacuum that includes a misalignment along the EW singlet.

The only simple, and physically relevant case, where the tadpole for $\varphi_0$ is absent, is when all pre-Yukawas are aligned with one of the two SU(2)$_R$ doublets, i.e. $y_{f2}=0$ (or $y_{f1}=0$) for which the minimum solution $\beta=0$ (or $\beta = \pi/2$) assures $Y_{f2}=0$ and the vanishing of all tadpoles. Note that for $\delta m_R = 0$ and aligned pre-Yukawas, i.e. when $\frac{y_{f2}}{y_{f1}}$ is a fermion-universal quantity, one can have the same situation. In fact, the rotation in Eq.~(\ref{eq:Omegabeta}) can be used to eliminate $\beta$ from all physical quantities and align the pre-Yukawas with only one of the two doublets. The vacuum is, therefore, simply parametrised by a single angle $\theta$.
The misalignment equation can then be easily solved
\begin{equation}
\cos \theta = \frac{16 B M/f^4 + 8 B \delta m_R/f^4}{2 C_t (|y_{t1}|^2 + |y_{b1}|^2) - C_g (3 g^2 + g'^2)}\,.
\end{equation}
This is the same result from the minimal case $\SU(4)/\Sp(4)$~\cite{Cacciapaglia:2014uja}. Furthermore, the second doublet decouples from the fermions and does not develop a VEV, thus it will act as an inert doublet.

%%%%%%%%%%%%%%%%%%%%%%%%%%%%%%%%%%%%%%
\subsection{Variations on the top mass and the vacuum misalignment}

In the previous subsection we have followed the simplifying assumption that the top and bottom masses come from bilinear couplings to the underlying techni-fermions. In practice, the key assumption is that the spurion generating the top and bottom masses is embedded in the global symmetries of the model in the same way as the Higgs bosons themselves. This scenario may be generated by an Extended Technicolor sector~\cite{Eichten:1979ah}.
Another possibility if offered by Bosonic Technicolor~\cite{Simmons:1988fu,Samuel:1990dq}, where the couplings between SM and techni-fermions are mediated by a scalar field, neutral under the FCD interactions. As the mediator couples to the techni-fermions via Yukawa-like interactions, in the simplest incarnation the spurion will have the same symmetry properties as the case we studied. Note that this case is also realised in the so-called models of half-composite Higgs~\cite{Antipin:2015jia,Agugliaro:2016clv}, where the mediator (which has the quantum numbers of the SM Higgs) is assumed to be as light as the composite states.
In the rest of this section we will explore how other mechanisms for the SM fermion mass generation may affect our conclusions on the vacuum misalignment.

A more recent scenario, which gained popularity in the past decade, is the mechanism of partial compositeness~\cite{Kaplan:1991dc}, where the elementary SM fermions mix directly with a spin-1/2 composite state. The masses of the SM fermions, thus, are generated after integrating out the heavy resonances. The symmetry properties of the spurions, now, depend on the representation of the global symmetry the composite operators belong to. From an effective field theory (EFT) point of view, there are infinite possible choices of increasing complexity and only phenomenological considerations may help preferring one over the other (see, for instance, the discussion in Ref.~\cite{Gripaios:2009pe}).
Instead, here we will rely on the indications spawning from concrete underlying models. This will not only fix the representation of the composite fermions, but also provide a controlled power counting for the construction of effective operators.

As a first example, we will follow the proposal from Ref.~\cite{Sannino:2016sfx}, where techni-scalars are added to the theory. The fermionic bound states are thus made of one fermion and one scalar, and the linear mixing of the SM fermions is added in terms of renormalisable Yukawa interactions. The main advantage of this class of models is that the interactions giving rise to fermion masses are always relevant without the need for large anomalous dimensions of the composite fermions. Moreover, hierarchies in the SM fermion masses can be easily obtained by choosing hierarchical Yukawa couplings or introducing large masses for the scalars (leading to multi-scale flavour generation~\cite{Cacciapaglia:2015dsa,Panico:2016ull}). The price to pay is that the underlying theory is not natural, as the masses of the techni-scalars remain unprotected: one possible solution might be that the theory becomes asymptotically safe at high energies~\cite{Pelaggi:2017wzr}. While we limit ourselves to fields that give mass to the top and bottom, additional scalars can be introduced to give mass to leptons and the light quarks without destabilising the theory~\cite{Sannino:2016sfx,Cacciapaglia:2017cdi,Sannino:2017utc}. The most minimal model that extends our $\SU(6)\to \Sp(6)$ model would thus contain a complex techni-scalar $\mathcal{S}_t$ transforming in the same representation as the techni-fermion under the FCD interactions, and that carries SM charges as indicated in Table~\ref{tab:TCscalars} (Model 1).
A prediction of this class of models is that the composite fermions, containing only one techni-fermion, always transform as the fundamental of the global symmetry $\SU(6)$. 
The Yukawa couplings in the underlying theory read:
\begin{equation}
- \mathcal{L}_{\rm Yuk}^{\rm FPC} =  \tilde{y}_{L}\ Q_L \mathcal{S}_t \psi_1 - \tilde{y}_{R1}\ t_R^c \mathcal{S}^\ast_t \psi_2^{(d)} -  \tilde{y}_{R2}\ t_R^c \mathcal{S}^\ast_t \psi_3^{(d)} + \tilde{y}_{bR1}\ b_R^c \mathcal{S}^\ast_t \psi_2^{(u)} +  \tilde{y}_{bR2}\ b_R^c \mathcal{S}^\ast_t \psi_3^{(u)} \,,
\end{equation}
where the indices $(u)$ and $(d)$ indicate the up and down components of the $\SU(2)_R$ anti-doublets $\psi_{2,3}$, and the appropriate gauge contractions are left understood.
A general operator analysis of the low energy EFT for this class of theories has been recently presented in Ref.~\cite{Cacciapaglia:2017cdi}.
It has been shown that, at leading order in the chiral expansion, operators corresponding to Eqs~\eqref{eq:topmass} and \eqref{eq:botmass} are generated, with
\begin{equation}
y_{t1/2} = \frac{1}{4 \pi}\ C_{\rm Yuk}\  \tilde{y}_L \tilde{y}_{R1/2}\,, \qquad y_{b1/2} = \frac{1}{4 \pi}\ C_{\rm Yuk}\  \tilde{y}_L \tilde{y}_{bR1/2}\,, 
\end{equation}
where $C_{\rm Yuk}$ is a form factor. An operator contributing to the vacuum misalignment can also be constructed at order $\tilde{y}_L^2 \tilde{y}_R^2$, and it matches with the form of Eq.~\eqref{eq:Vtop}.  This matching for both effective Yukawa and potential can be easily explained by the observation that the anti-symmetric spurions we used in the previous section (given in Appendix~\ref{app:topproj}) can be constructed starting from the spurions in the fundamental that we use to describe partial compositeness in this model~\cite{Alanne:2018wtp}. 
This model, however, contains new contributions to the potential in the form of operators at order $\tilde{y}_L^4$ and $\tilde{y}_R^4$~\cite{Cacciapaglia:2017cdi,Alanne:2018wtp}: we checked that while the one proportional to $\tilde{y}_R^4$ vanishes, the one of order $\tilde{y}_L^4$ generates a potential term proportional to $c_\theta^2$ but it only contains a tadpole for the would-be Higgs $h_1$. This means that, while the spectrum will receive additional contributions from this operator, the discussion on the vacuum alignment remains unaffected.

\begin{table}[tb]
\begin{tabular}{|l|c|ccc|c|}
\hline
 & $G_{\rm TC}$ & $\SU(3)_c$ & $\SU(2)_L$ & $\U(1)_Y$ & global sym.\\
\hline
\multicolumn{6}{c}{\phantom{xxx} Model 1: $\Sp(2)_{\rm TC}$ with techni-scalars, $\psi = \tiny\yng(1)$ \phantom{xxx}} \\
\hline
$\mathcal{S}_t$ & $\tiny\yng(1)$ & $\bar 3$ & $1$ & $-1/6$  & $\SU(6)_\psi \times \Sp(6)_\mathcal{S}$ \\ 
\hline
\multicolumn{6}{c}{\phantom{xxx} Model 2: $\Sp(2N)_{\rm TC}$ with techni-fermions, $\psi = \tiny\yng(1)$ \phantom{xxx}} \\
\hline
$\chi$ & \multirow{2}{*}{$\tiny\yng(1,1)$} & $3$ & $1$ & $2/3$  & \multirow{2}{*}{$\SU(6)_\psi \times \SU(6)_\chi \times \U(1)$ }\\ 
$\tilde{\chi}$ &  & $\bar 3$ & $1$ & $-2/3$  & \\ 
\hline
\multicolumn{6}{c}{\phantom{xxx} Model 3: $\SO(11)_{\rm TC}$ or $\SO(13)_{\rm TC}$ with techni-fermions, $\psi = {\bf Spin}$ \phantom{xxx}} \\
\hline
$\chi$ & \multirow{2}{*}{$\tiny\yng(1)$} & $3$ & $1$ & $2/3$  & \multirow{2}{*}{$\SU(6)_\psi \times \SU(6)_\chi \times \U(1)$ }\\ 
$\tilde{\chi}$ &  & $\bar 3$ & $1$ & $-2/3$  & \\ 
\hline
\end{tabular}
\caption{FCD models that generate partial compositeness in the confined phase. In Model 1, techni-scalars are added in the same representation as the techni-fermions (specifically for $\Sp(2)$). In Models 2 and 3, the composite fermions are made of 2 $\psi$'s and an additional fermion $\chi$ in a different FCD representation. In all models, the component of $\psi$ transform under the SM interactions as in Table~\ref{tab:models}.} \label{tab:TCscalars}
\end{table}

Another possibility to construct  underlying theories of partial compositeness is to assume that the fermionic bound states are made of 3 techni-fermions. \footnote{In principle, techni-fermion/techni-gluon bound states as well as multi-fermion ones may form. For the former the problem is that the techni-fermions need to be in the adjoint representation, thus their multiplicity is strongly limited by asymptotic freedom. In the latter case, the linear mixing with the SM fermions would arise at dimension larger than 9, implying a stronger suppression.}  In Ref.~\cite{Ferretti:2013kya} it has been shown that, for techni-fermions in a pseudo-real representation, the only possibility is to add at least one other species of techni-fermions, $\chi$, in a different representation of the FCD gauge symmetry. The only two cases allowed by asymptotic freedom are listed in Table~\ref{tab:TCscalars}. The model based on $\Sp(2N)_{\rm TC}$ has been first proposed in Ref.~\cite{Barnard:2013zea}. Interestingly, the two models would be indistinguishable in an EFT approach, except for the properties of the singlet mesons associated to the $\U(1)$ global symmetry~\cite{Belyaev:2015hgo,Belyaev:2016ftv} and, of course, by the mass spectra~\cite{Bizot:2016zyu,Bennett:2017kga}.
A big limitation of this approach is that the multiplicity of the top partners is strongly limited by the loss of asymptotic freedom \footnote{By increasing the number of fermions, before losing asymptotic freedom, the model will enter the conformal window~\cite{Sannino:2009aw,Ryttov:2009yw}, thus losing the formation of a condensate and mass gap. This further limits the Model 2 to $N_c = 2$~\cite{Ferretti:2016upr} for the minimal fermion content in Table~\ref{tab:TCscalars}.}, so that it is unlikely to be able to give mass to all SM fermions via partial compositeness. In this class of models, the representation of the composite fermions depends on their fermion content. Being always made of two techni-fermions $\psi$ and one $\chi$, they can only belong to the following representations of $\SU(6)$:
\begin{equation}
{\tiny\yng(1,1)} = {\bf 15}_{\SU(6)}\,, \quad {\tiny\yng(2)} = {\bf 21}_{\SU(6)}\,, \quad {\tiny\yng(2,1,1,1,1)} = {\bf 35}_{\SU(6)}\,,
\end{equation}
i.e. all possible 2-index representations. In the following we will only consider the symmetric $\bf 21$, as it's the only case where operators with two pre-Yukawa \footnote{The pre-Yukawas are couplings generating the linear mixing of the SM fermions with the composite fermions. They are 4-fermion interactions in the underlying theory, like in the case we considered in the previous section, but only linear in the SM fields.} insertions are forbidden. In an EFT analysis, the pre-Yukawas are spurions that carry power counting in the derivative expansion, thus operators with larger number of spurion insertions are higher order compared to ones containing a smaller number of spurions. Their coefficient is thus expected to be naturally suppressed.
Operators with two pre-Yukawa insertions appear at the lowest possible order and they are expected to give too large contributions to the pNGB masses unless they are fine-tuned (see, for instance, Refs~\cite{Csaki:2017cep,Alanne:2018wtp}). One possible way to fine-tune them is to assume that they are dominantly generated by loops of the composite fermions~\cite{Matsedonskyi:2012ym,Marzocca:2012zn}, which are assumed to be light and weakly coupled to the pNGBs~\cite{Contino:2011np}. We will follow, here, the complementary approach of selecting a case where the operators are naturally appearing at higher order, leading to natural choices for the form factors. This is the case of the symmetric representation of $\Sp(6)$, for which operators with two spurion insertions are forbidden~\cite{Alanne:2018wtp}, and the operators only arise at the order of four spurion insertions. The symmetric representation decomposes, under the symmetry $\SU(2)_L \times \SU(2)_{R1} \times \SU(2)_{R2}$ as:
\begin{equation}
{\bf 21}_{\rm SU(6)} = (2,2,1) \oplus (2,1,2) \oplus (1,2, 2) \oplus (3,1,1) \oplus (1,3,1) \oplus (1,1,3)\,.
\end{equation}
The left-handed quarks can couple linearly with either one of the $\SU(2)_L$ doublets, thus allowing for two pre-Yukawa couplings $y_{L1/2}$. On the other hand, for the right-handed top, there are four possible choices: either one of the two $\SU(2)_R$ triplets, or the two neutral components of the $(1,2,2)$ bi-doublet. In the following, we will choose to embed it in both the $\SU(2)_R$ triplets: the explicit form of the spurions $S_L$ and $S_R$, which are a generalisation of the ones discussed in Ref.~\cite{Alanne:2018wtp} for the minimal $\SU(4)/\Sp(4)$ model, are shown in Appendix~\ref{app:topproj}.
The top mass is generated by an operator in the form~\cite{Golterman:2017vdj,Alanne:2018wtp}:
\begin{equation}
 - \frac{C_{\rm t} f}{4 \pi} \mbox{Tr} [S_L \cdot \Sigma^\dagger \cdot S_R \cdot \Sigma^\dagger] = - Y_{t1} f s_{2\theta} + \dots
\end{equation}
where 
\begin{equation}
y_{t1/2} = \frac{1}{4 \pi}\ C_{\rm t}\  y_{L1/2} y_{R}\,.
\end{equation}
Thus, the top mass has a different $\theta$--dependence from the case of bi-linear operators. We further observe that the linear couplings (which are included in the dots) have the same structure as the ones in Eq.\eqref{eq:topmass}, except for a different $\theta$--dependence and the appearance of a couplings of $\eta_1$ proportional to the top mass. An operator that generates a potential for the misalignment angle can be constructed following Ref.~\cite{Alanne:2018wtp}. We have checked that tadpoles for the non-Higgs pNGBs are generated by the operators proportional to $y_L^4$ and $y_L^2 y_R^2$, however they have a similar structure as Eq.\eqref{eq:Vtop}, except for a different dependence on $\theta$. \footnote{The operator of order $y_L^4$, which does not contain $y_R$, has an overall normalisation factor $Y_{t1}^2/y_R^4$.} Thus, for this case too, our considerations relative to the vacuum misalignment stay unchanged.  The spectrum of the pNGBs would, however, be different.
We should remark that a different choice for the $t_R$ spurion would change the dependence on $\beta$ of the potential.

We thus conclude that our conclusions about the vacuum misalignment, summarised at the end of Section~\ref{sec:Yuktopbot}, are solid and not specific to the mechanism generating the top and bottom masses we chose in the previous section. 
We, in fact, provide examples of similar misalignment potential generated for models with partial compositeness for the top, and ones bases on half-composite Higgs.

%%%%%%%%%%%%%%%%%%%%%%%%%%%%%%%%%%%%%%
\subsection{A composite inert 2HDM} \label{sec:i2HDM}

Before concluding the section, we would like to further investigate the aligned case that leads to a composite inert 2HDM.
We recall that this situation is achieved when all fermions couple to the same $\SU(2)_R$ doublet (thus, we can choose $y_{f2}=0$ without loss of generality), or when the top and bottom Yukawas are aligned and $m_{R1} = m_{R2}$ (in which case, the Yukawas can be rotated on the first doublet for free).
The minimum is thus characterised by
\begin{equation}
\beta = 0\,, \;\; \Rightarrow \;\; Y_{f2}=0\,, \;\; Y_{f1}=y_{f1}\,.
\end{equation}
As already mentioned, the minimisation of the vacuum imposes a relation between the angle $\theta$ and the average mass $M$ very similar to the one in the $\SU(4)/\Sp(4)$ case; similarly, we find that the would-be Higgs boson $h_1$ does not mix to the other CP-even scalars $h_2$ and $\varphi_0$ and its mass and couplings follow the same formulas as in the $\SU(4)/\Sp(4)$ case~\cite{Cacciapaglia:2014uja}: 
\begin{equation} \label{eq:mh0}
m_{h_1}^2 = \frac{C_t}{4} m_t^2 - \frac{C_g}{16} (2 m_W^2 + m_Z^2)\,, \quad \frac{g_{hXX}}{g_{hXX}^{\rm SM}} = c_\theta\,.
\end{equation}
The Higgs physics will, therefore, remain the same as in the minimal case~\cite{Arbey:2015exa}. In particular, we remind the reader that the strong coefficient $C_t \sim 2$ (which is calculable once the dynamics is fixed) needs to be tuned to reproduce the correct Higgs mass of $125$~GeV.

The 10 pNGB, in addition to the SM-like Higgs and exact Goldstones, show peculiar properties. First we notice that the $\SU(2)_{R2}$ acting on the second doublet is not broken either by the techni-fermion mass or by the Yukawa couplings, however it's broken to $U(1)$ by the partial gauging of hypercharge.
Thus, within the minimal set of spurions, we can identify an unbroken $U(1)$ under which some of the PNGBs are charged and stable (being the elementary SM fields uncharged). The presence of this unbroken symmetry is guaranteed as long as the only spurions in the models are the ones described above, i.e. gauging of the electroweak symmetry, techni-fermion masses and Yukawa couplings (involving only the first Higgs doublet). Higher order operators containing such spurions will preserve the symmetry at all orders. If additional spurions are added, the symmetry is preserved as long as they are invariant under $\SU(2)_{R2}$. Thus, the presence of the $U(1)$ is a rather solid.

If we defined the dark matter (DM) charge $Q_{\rm DM} = 2 T^3_{R2}$, analysing the pNGB structure in Eq.(\ref{eq:pNGBmatrix}), we see that the following fields have charge $Q_{\rm DM} = 1$:
\begin{equation} 
H^0 = \frac{h_2 - i A_0}{\sqrt{2}}\,, \; \eta^0 = \frac{\varphi_0 - i \eta_3}{\sqrt{2}}\,, \; H^+\,, \; \eta^+\,.
\end{equation}
The two additional singlets, $\eta_1$ and $\eta_2$, are not charged under $\U(1)_{\rm DM}$.

Both the DM-charged and neutral pairs of states feature a mixing: to calculate the spectrum, we solve the minimum condition in terms of the average mass $M$, and leave the value of the minimum angle $\theta$ as a free parameter.  
The resulting mass in the DM-charged neutral sector, in the basis $(H^0, \eta^0)$, reads
\begin{equation} \label{eq:M2neut}
\mathcal{M}^2_{\rm neut.} = \frac{C_t f^2}{8} \left( \begin{array}{cc}
Y_{t1}^2 (1+c_\theta) - 8 K_\delta & Y_{t1}^2 s_\theta \\
Y_{t1}^2 s_\theta & Y_{t1}^2 (1+ c_\theta -2\Delta c_\theta) - 4 (1-\Delta) K_\delta - \frac{C_g}{C_t} (3 g^2 + {g'}^2) (1-\Delta) c_\theta
\end{array} \right)\,,
\end{equation}
where, for convenience of notation, we have defined 
\begin{equation}
K_\delta = \frac{2 B \delta m_R}{C_t f^4}\,. 
\end{equation}
Note that, as expected, the mixing depends linearly on the EW symmetry breaking parameter $s_\theta$, so it is suppressed at small misalignment angles.
The system of the two charged states $(H^\pm, \eta^\pm)$ have an additional contribution from the gauging of hypercharge, which is thus responsible for generating a small mass splitting between them:
\begin{equation} \label{eq:M2charg}
 \mathcal{M}^2_{\rm charg.} = \mathcal{M}^2_{\rm neut.} + 
 \frac{C_g {g'}^2 f^2}{4} \left( \begin{array}{cc}
(1-c_\theta) & 0 \\
0 & (1+c_\theta)
\end{array} \right)\,.
\end{equation}
As $C_g > 0$, the charged states will always be slightly heavier than the neutral ones.
Finally, for the neutral singlets in the basis $(\eta_1, \eta_2)$, the mass matrix reads
\begin{equation}
 \mathcal{M}^2_{\eta} = \frac{m_h^2}{s_\theta^2}  \begin{pmatrix}
1 & \frac{1}{\sqrt{3}}\Delta c_\theta\\
 \frac{1}{\sqrt{3}}\Delta c_\theta & \frac{1}{3}  (2 (1-\Delta) + c_\theta) c_\theta \end{pmatrix} - C_t f^2  K_\delta\ \begin{pmatrix}
 0 & \frac{1+\Delta}{2 \sqrt{3}} \\
\frac{1+\Delta}{2 \sqrt{3}} & \frac{3-\Delta}{3} \end{pmatrix} \,,
\end{equation}
where we have used the relation between $C_t$ and the Higgs mass to simplify the expression.

\begin{figure}[tb!]
\begin{center}
\includegraphics[width=12cm]{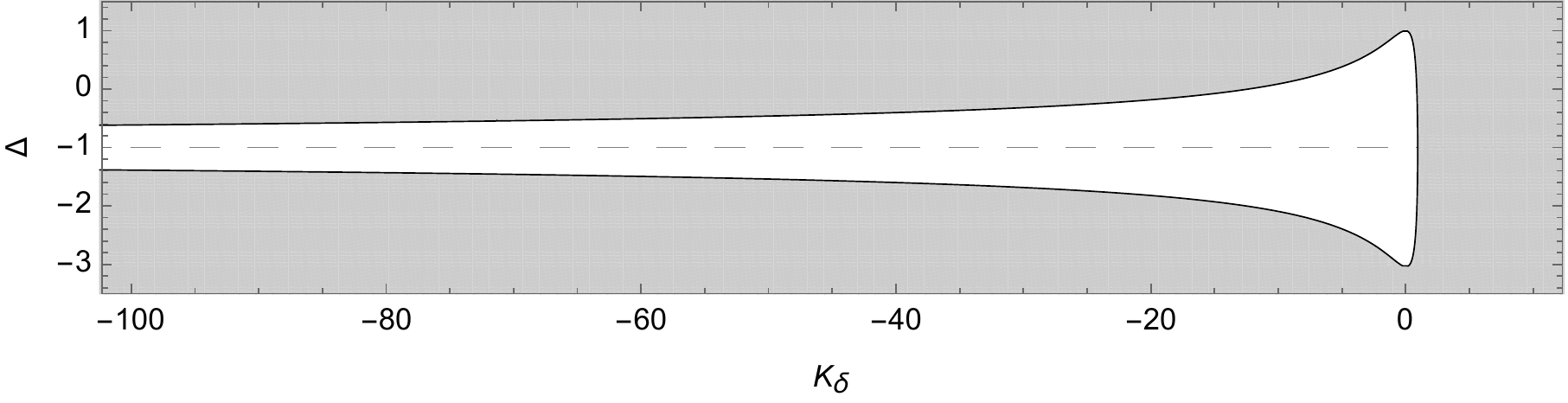}
\end{center}
\caption{Allowed parameter space as a function of $\Delta$ and $K_\delta$ for $\theta = 0.2$.} \label{fig:paramspace}
\end{figure}

We remark that, once $C_t$ is fixed to reproduce the measured Higgs mass, the model has 3 free parameters: $\Delta$, $K_\delta$ and $C_g$. The misalignment angle $\theta$ is also not fixed, but provides a direct relation between the EW scale and the condensation scale. However, once we fix a small $\theta$, not all values of the other parameters are physical. This can be seen from the fact that the mass matrices, in particular the one for the singlets $\eta_{1/2}$ that does not depend on $C_g$, develops negative eigenstates. This signals that the vacuum has not been properly chosen and, as the fields are singlets of the EW symmetry, it indicates that the EW preserving vacuum should be modified.
In Fig.~\ref{fig:paramspace} we show the allowed parameter space for $\theta = 0.2$ (corresponding to $2\sqrt{2} f = 1.2$~TeV), which features a characteristic nail-shape, roughly symmetric around $\Delta=-1$. The shape and size of the allowed parameter space depend only mildly on the misalignment angle. From the definition of $\Delta$, in Eq.(\ref{eq:MDelta}), we see that its value is bound between $1$ and $-1$ for positive masses, with $-1$ corresponding to $m_R \gg m_L$. Thus, values $\Delta < -1$ can only be reached for $m_R < -m_L$. On the other hand, large negative values for $K_\delta$ are possible for large values of the masses $m_{R1}$ and $m_{R2}$: the throat, therefore, can be consistently achieved for $m_{R2}$ very large (for $\Delta > -1$) or $m_{R1}$ large and negative (for $\Delta < -1$). In both cases, one of the three underlying fermions becomes very heavy and thus decouples, invalidating the effective description of the model. For this reason, in the following, we will focus on the head of the nail.

\begin{figure}[tb!]
\begin{center}
\includegraphics[width=6.5cm]{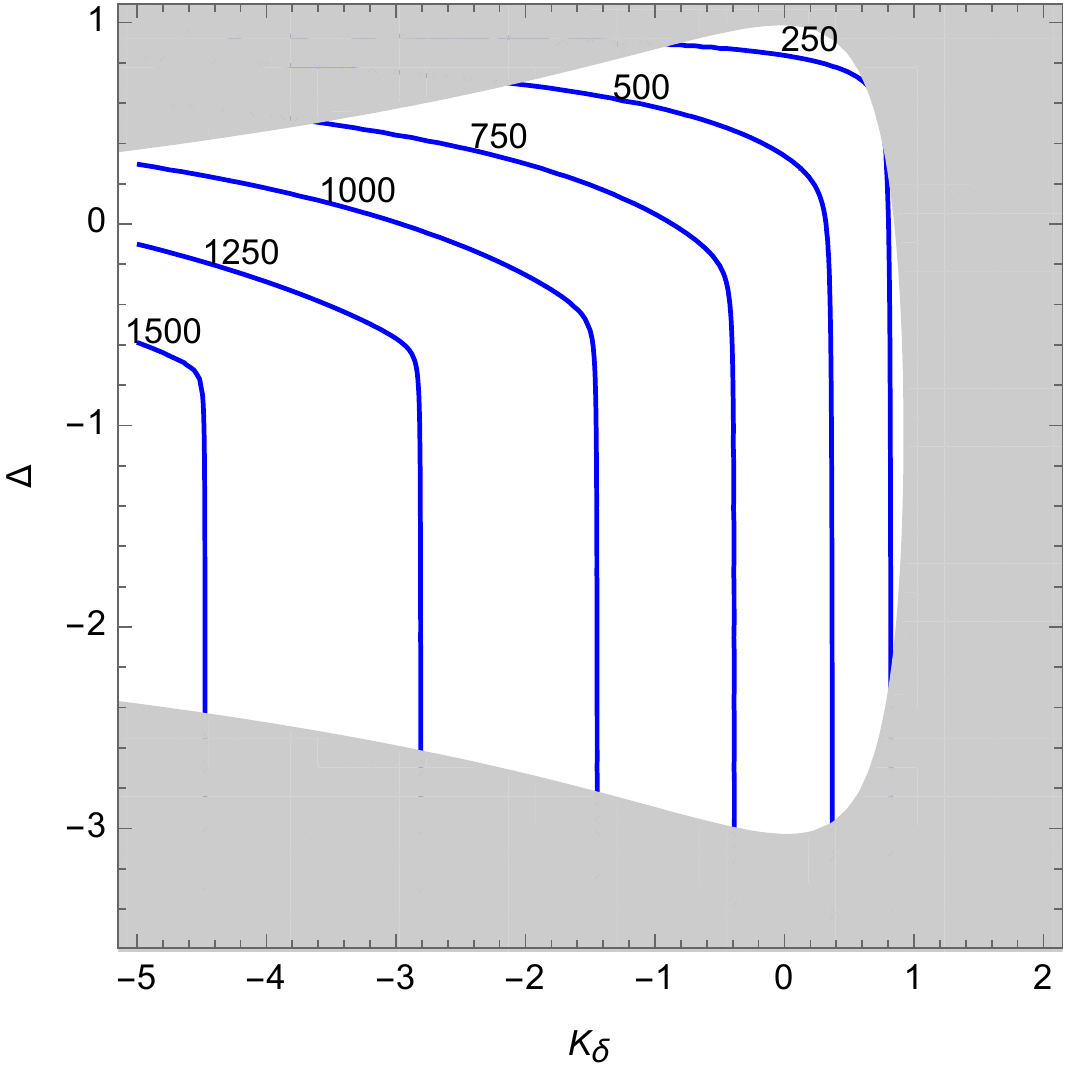} \hspace{1cm}
\includegraphics[width=6.5cm]{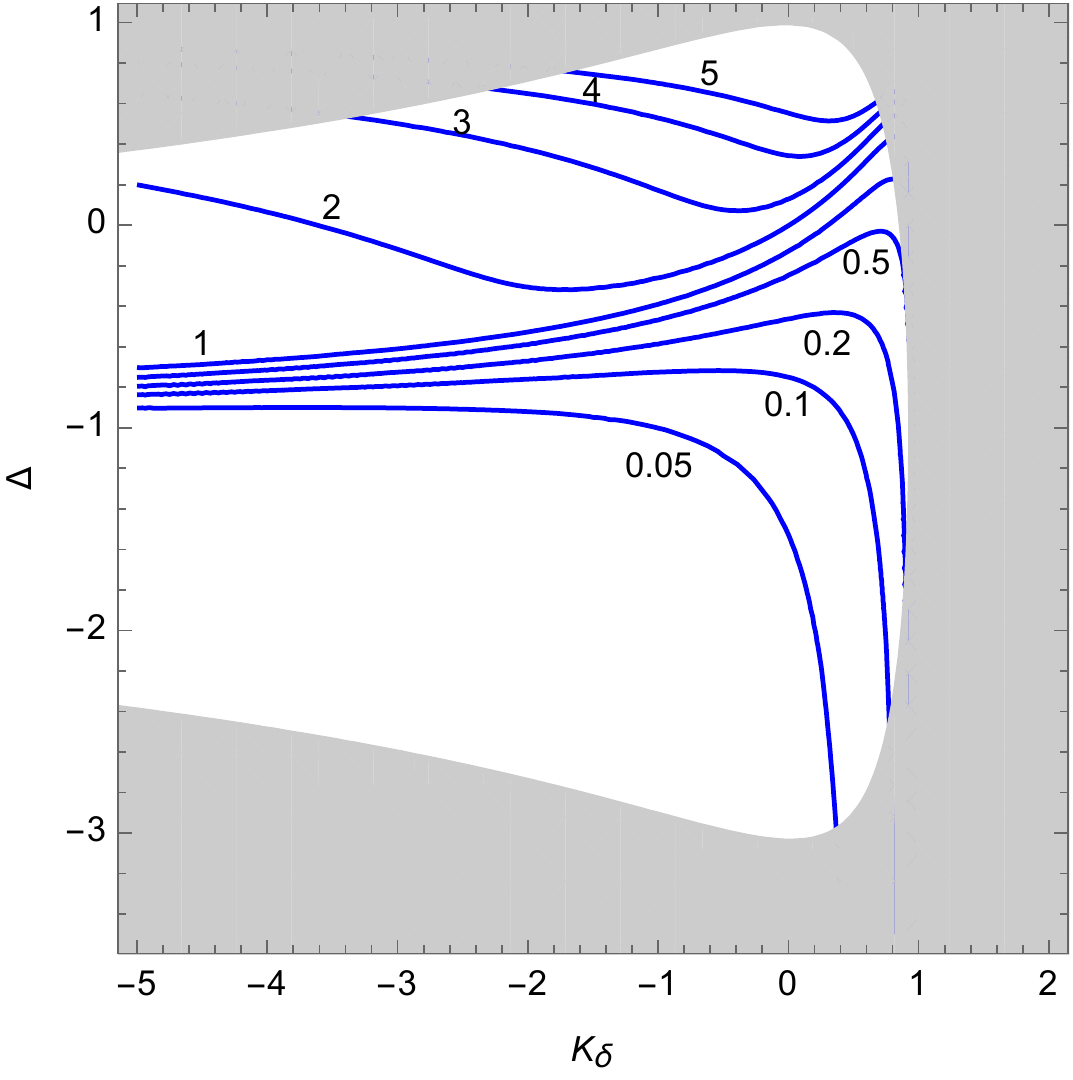}
\end{center}
\caption{Contours, as a function of $\Delta$ and $K_\delta$, of the mass of the Dark Matter candidate (left) and the mass splitting with the nearby charged scalar (right). All values are in GeV, and we fixed $C_g = 1/3\ C_t$ and $\theta = 0.2$ (i.e. $2 \sqrt{2} f=1.2$~TeV). The grey area is theoretically unaccessible.} \label{fig:contoursi2HDM}
\end{figure}

We now present the mass of the lightest stable DM-charged scalar (which is neutral) in the allowed parameter space, as shown in the left panel of Fig.\ref{fig:contoursi2HDM}, where we fix $\theta=0.2$ and $C_g = 1/3\ C_t$ \footnote{The factor of $1/3$ corresponds to the colour multiplicity that is present for top loops, but not for gauge boson loops. This is just an estimate, the impact of gauge loops to the spectrum being marginal.}.
We remark that below a critical value of $\Delta$, that depends on $K_\delta$, the mass of the lightest state becomes independent on $\Delta$: this behaviour can be traced back in the form of the mass matrix in Eq.(\ref{eq:M2neut}), as the entry corresponding to the $\SU(2)_L$ doublet does not depend on $\Delta$. Thus, the region with vertical lines corresponds to the lightest mass eigenstate dominantly made of the doublet. Note that this region is very likely to be excluded by direct detection, if this state saturates the relic abundance~\cite{Hambye:2009pw}. This behaviour is matched by the mass difference between the lightest charged and neutral states, which is shown in the right panel of Fig.\ref{fig:contoursi2HDM}. We see that a sub-GeV mass splitting is observed in the region where the lightest neutral state belongs to the doublet, in agreement with the mass matrix in Eq.(\ref{eq:M2charg}). We can therefore conclude that the interesting parameter space for Dark Matter will correspond to values of $\Delta$ larger than $-1$, which is also consistent with our initial assumption of positive masses.
\begin{figure}[tb!]
\begin{center}
\includegraphics[width=6.5cm]{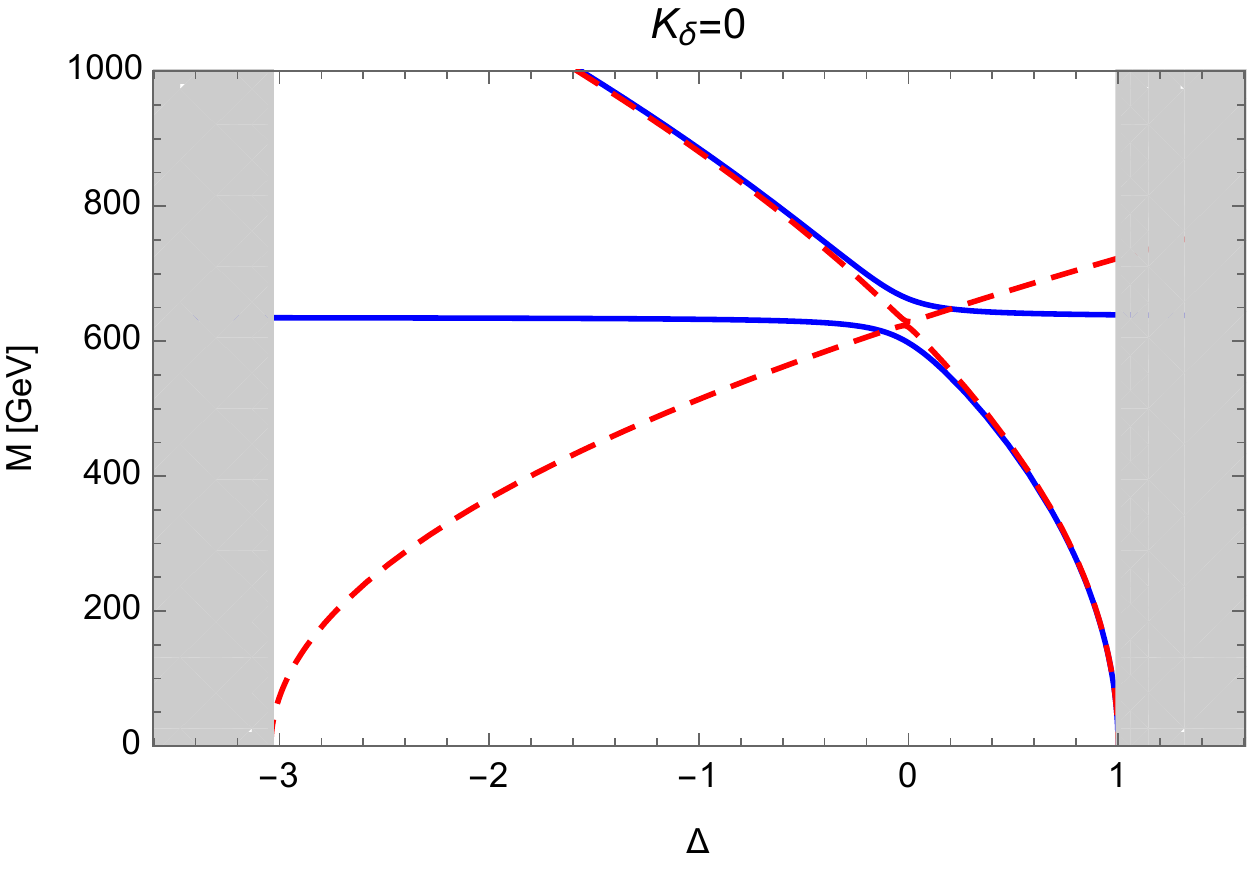} \hspace{1cm}
\includegraphics[width=6.5cm]{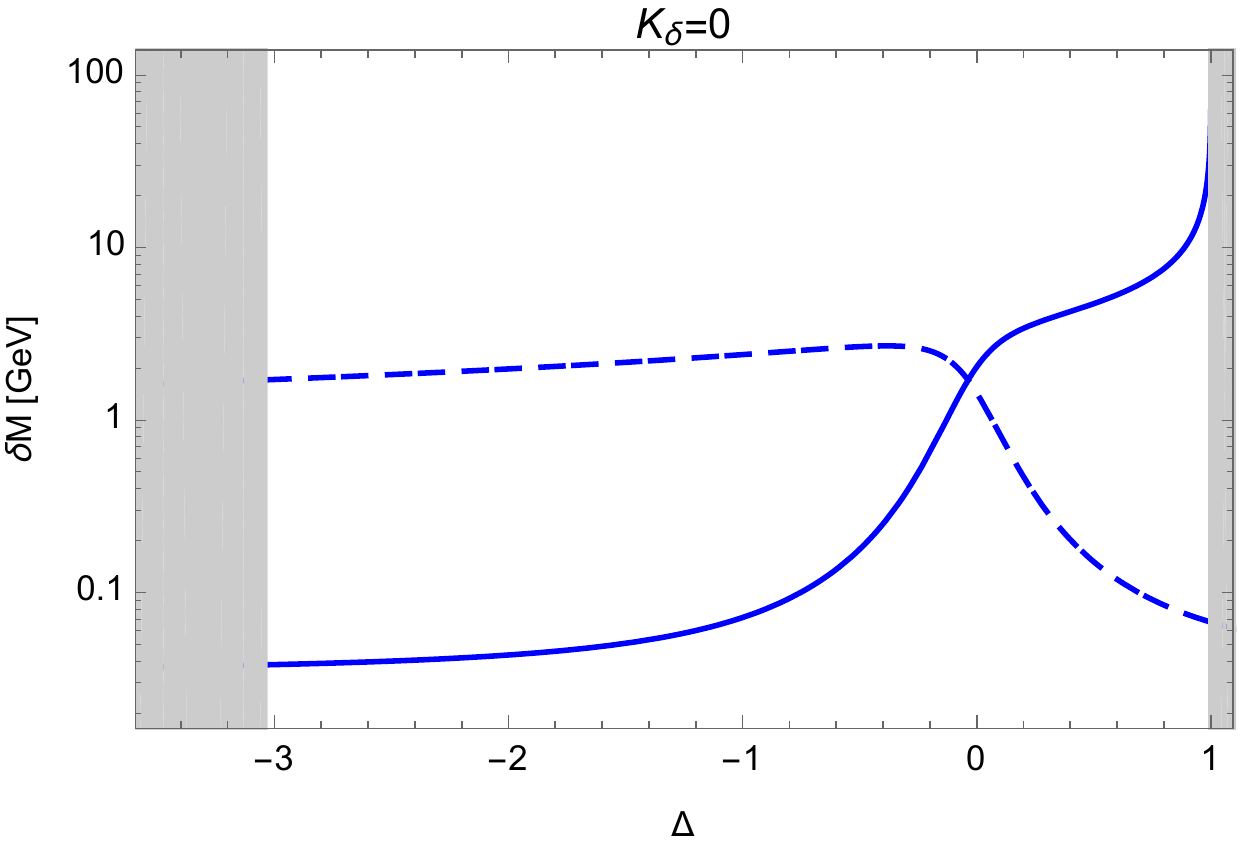}
\includegraphics[width=6.5cm]{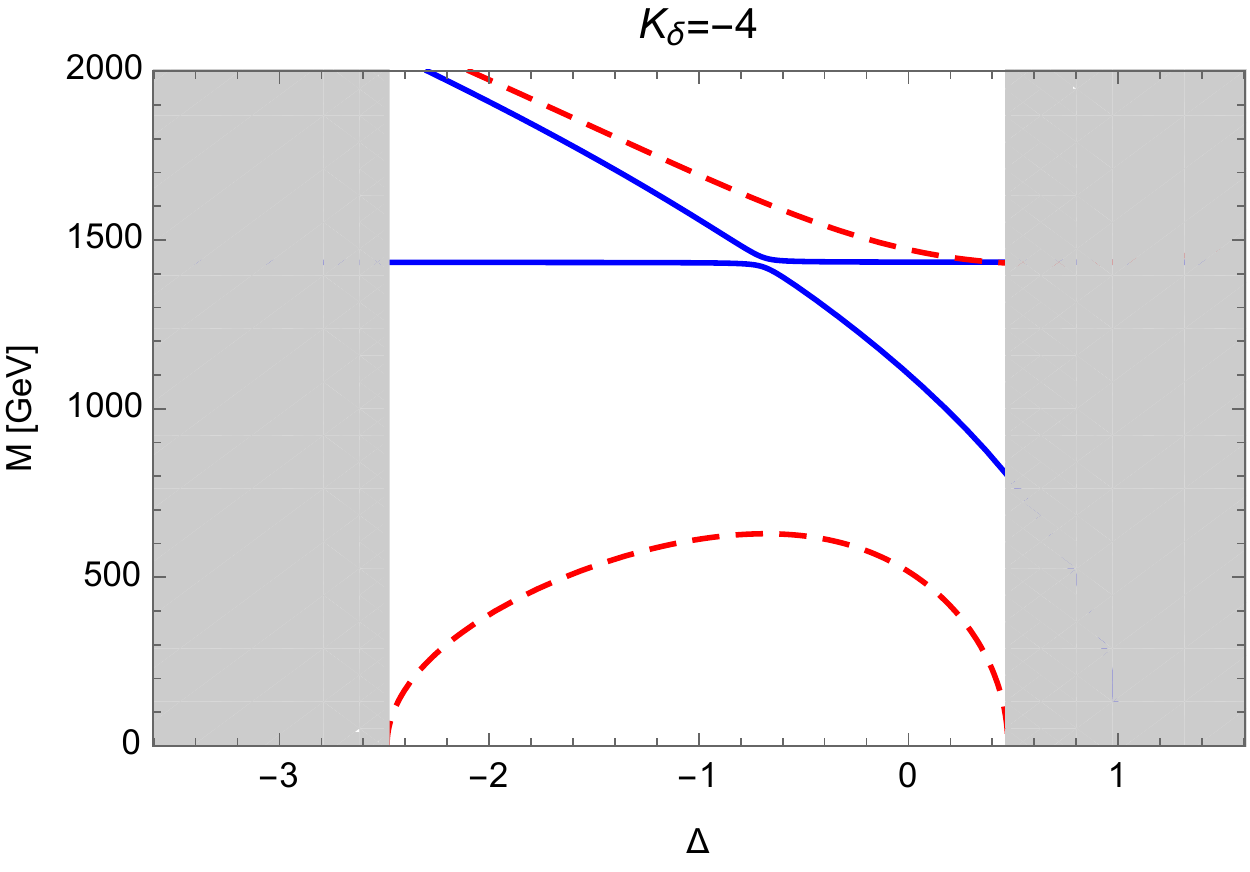} \hspace{1cm}
\includegraphics[width=6.5cm]{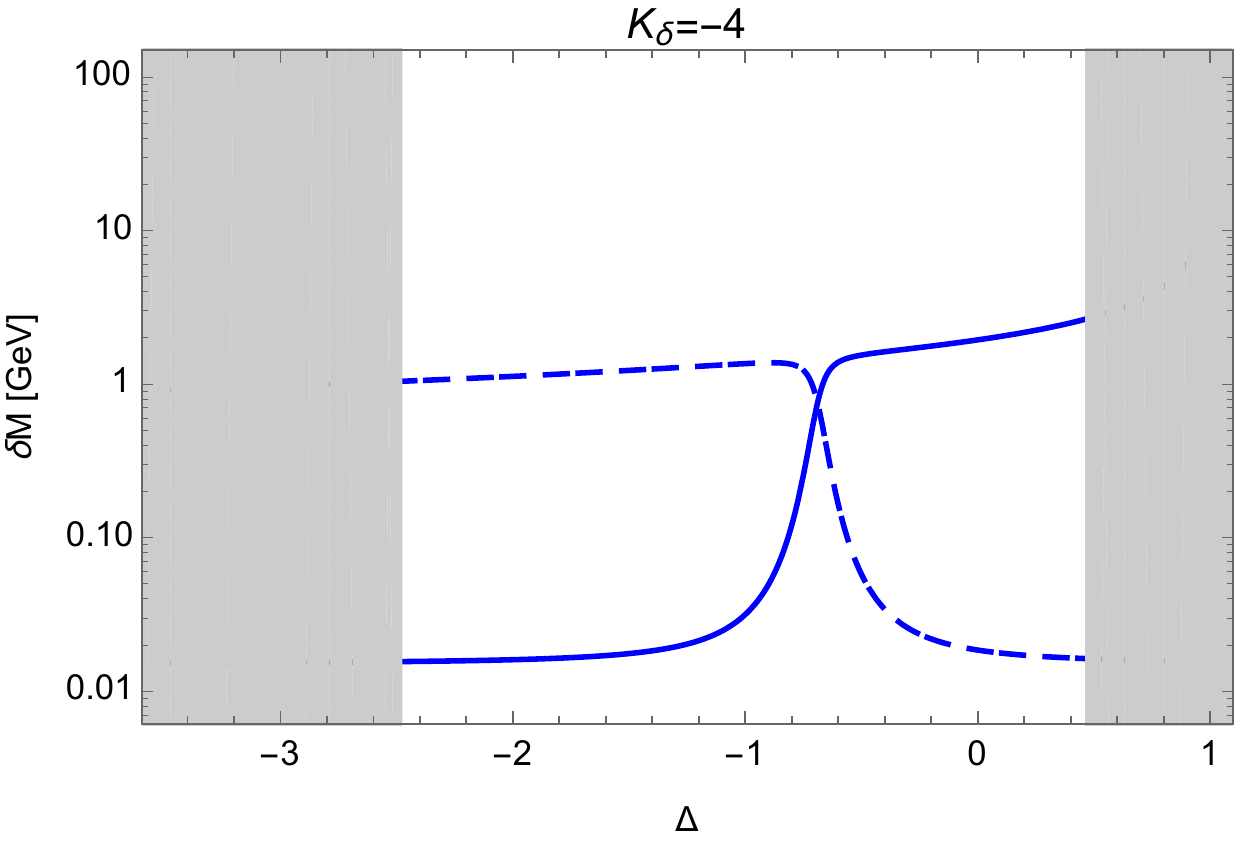}
\end{center}
\caption{(Left) Masses of the even (dashed red) and odd (blue) scalars as a function of $\Delta$. (Right) Mass splitting between the charged and neutral scalars in the two odd tears, with the light one in solid lines. The two rows correspond to $K_\delta = 0.$ and $-4$ respectively. All values are in GeV, and we fixed $C_g = 1/3\ C_t$ and $\theta = 0.2$ (i.e., $2 \sqrt{2} f=1.2$~TeV). The grey areas are theoretically unaccessible.} \label{fig:spectrumi2HDM}
\end{figure}
To further clarify this behaviour, in Fig.~\ref{fig:spectrumi2HDM} we slice the parameter space for fixed values of $K_\delta = 0$ (upper row) and $K_\delta = -4$ (lower row). It can be clearly seen the level crossing point between the two odd states (blue lines), which corresponds to an switch of the mass differences (right plots). Interestingly, we see that there is always a pair of states with a mass difference below a GeV, except very close to the crossing region. This may lead to long-lived charged particles at the LHC. In the left plots, we also show in red the eigenvalues for the singlet pseudo-scalars. The plot shows the presence of a light state, whose mass is always below $m_h/s_\theta \sim 600$~GeV, which becomes tachyonic close to the unaccessible region.

As already mentioned, the $\SU(2)_L$ doublet cannot be a thermal DM candidate because of the coupling to the $Z$ boson. However, in this model the singlet $\eta^0$ also features a coupling to the $Z$ boson due to non-linear effects of the misalignment, as shown in Appendix~\ref{app:pNGBcoupl}. The coupling of the singlet is proportional to $(1-c_\theta)$, thus for small $\theta$ it arises at order $s_\theta^2$, as expected. Integrating out the $Z$ boson, this coupling will generate a coupling of the DM candidate with a vector quark current, which contributes to the spin-independent scattering cross section of the DM off nucleons. Following Ref.~\cite{Yu:2011by}, we estimate the cross sections off a nucleon $N = p,n$ to be
\begin{equation}
\sigma_{V, \eta^0 N} = \frac{(1-c_\theta)^2 g^4 m_N^2}{16 \pi c_W^4 m_Z^4} \times \left\{ \begin{array}{l}
\left( \frac{1}{4} - s_W^2 \right)^2\,, \qquad \mbox{for protons}, \;\; N=p\,; \\
\left( \frac{1}{4}\right)^2\,, \qquad \mbox{for neutrons}, \;\; N=n\,. \end{array} \right.
\end{equation}
Numerically, this leads to
\begin{equation}
\sigma_{V,\eta^0 p} \sim 2.7 \cdot 10^{-41}\; (1-c_\theta)^2\; \mbox{cm}^{2}\,, \qquad \sigma_{V,\eta^0 n} \sim 2.3 \cdot 10^{-39}\; (1-c_\theta)^2\; \mbox{cm}^{2}\,,
\end{equation}
where the smallness of the proton cross section is due to a cancellation in the Weinberg angle dependent factor.
By comparing with the strongest upper bound from direct detection from XENON 1T~\cite{Aprile:2018dbl} (see also LUX~\cite{Akerib:2016vxi} and PandaX-II~\cite{Cui:2017nnn} results), which ranges between $4 \cdot 10^{-47}\, \mbox{cm}^2$ (for $m_{\rm DM} = 30$~GeV) and $9 \cdot 10^{-46}\ \mbox{cm}^2$ (for $m_{\rm DM} = 1$~TeV), we see that the scattering to neutrons would require $s_\theta \sim 10^{-2}$ in order to evade the bound, thus rendering the fine tuning very severe. One possible way out would be to consider very low masses in the few GeV range, where the direct detection experiments loose sensitivity, however this scenario would be excluded by indirect detection, like for instance Fermi-LAT observation of photons from globular galaxies~\cite{Ahnen:2016qkx}.
Note finally that the mixing between the two states only makes the situation worse, as the couplings of the lightest eigenstate would be proportional to $(1-c_{2\alpha} c_\theta)$, $\alpha$ being the mixing angle between $\eta^0$ and $H^0$. The model also contains a coupling of $\eta^0$ to the Higgs and a direct coupling to fermions, however we checked that they give spin-independent cross sections below the exclusion limits.

One possible way out from this issue is to consider the case where the thermal relic abundance is very small, but an asymmetry in the DM number, generated during the phase transition at the condensation of the strong dynamics, saturates the relic abundance. In this case, no indirect detection bounds apply and low mass DM may be able to avoid the strong constraints from the $Z$-mediated scattering. Also, if the thermal relic density of the stable pNGB is smaller than the measured one, direct detection bounds can be avoided at the price of the scalar being only a minor component of Dark Matter. An extension of the model that contains other Dark Matter candidates would be necessary in such case.
We leave a more detailed study of the phenomenology of the Dark Matter candidate and of the other pNGBs for future investigation. In the next section, we will analyse a more general vacuum structure of the model.

%%%%%%%%%%%%%%%%%%%%%%%%%%%%%%%%%%%%%%%
\section{Most general (real) vacuum alignment} \label{sec:vacuum}

As we have seen in the previous section, unless the Yukawa couplings of the top and bottom quarks are aligned, a misalignment along a direction corresponding to a singlet is forced by the quark loops. In this section, we will therefore study this vacuum and its effects on the composite Higgs physics.
For simplicity, we will neglect the effect from the bottom quarks, which we assume to be numerically small, and focus on the top, gauge and techni-fermion mass contributions.

%%%%%%%%%%%%%%%%%%%%%%%%%
\subsection{Defining the $\gamma$-vacuum}

The new vacuum is defined in terms of three parameters: besides the two angles identified with the Higgs directions $X^4$ and $X^8$, a misalignment along the singlet associated to the generator $X^{13}$ is activated. The new vacuum is thus defined in terms of an $\SU(6)$ rotation that depends on three angles:
\begin{equation}
\Sigma_{\theta, \beta, \gamma} = \Omega (\theta, \beta, \gamma) \cdot \Sigma_0 \cdot \Omega^\dagger (\theta, \beta, \gamma)\,.
\end{equation}
The rotation can be written as
\begin{equation} \label{eq:Omegatbg}
\Omega (\theta, \beta, \gamma) = R_\beta \cdot R_\gamma \cdot \Omega_\theta \cdot R_\gamma^\dagger \cdot R_\beta^\dagger\,,
\end{equation}
where $R_\beta$ and $\Omega_\theta$ are defined in the previous Section in Eqs (\ref{eq:Omega}) and (\ref{eq:Omegabeta}), and
\begin{equation}
R_\gamma = e^{- i 2 \sqrt{2} \gamma\ S^{14}} = \begin{pmatrix}
\cos{\gamma}\ \mathbb{I}_2 & 0 & \sin{\gamma}\ \sigma^1\\
0 &\mathbb{I}_2 & 0 \\
-\sin{\gamma}\ \sigma^1 & 0 &  \cos{\gamma}\ \mathbb{I}_2
\end{pmatrix}\,.
\end{equation}
This property is due to the fact that the two sets of three generators $\{ X^4,\ X^{13},\ S^{14}\}$ and $\{ X^4,\ X^8,\ S^{21}\}$ form two partly-broken $\SU(2)$ subgroups of $\SU(6)$ (more details on the derivation of the above expression can be found in Appendix~\ref{app:genvac}).
As both $R_\beta$ and $R_\gamma$ are defined in term of unbroken generators around the vacuum $\Sigma_0$, the pNGBs can be described similarly to Eq.(\ref{eq:RbetaRot}), as:
\begin{equation}
\Sigma (\phi) = R_\beta \cdot R_\gamma \cdot \Omega_\theta \cdot U'_0 (\phi) \cdot \Sigma_0 \cdot \Omega_\theta^T \cdot R_\gamma^T \cdot R_\beta^T\,. 
\end{equation}
The role of $\beta$ in this vacuum is very similar to the one in the previous case (with $\gamma=0$), as it can be projected on the spurions.
The same can be done, in principle, for $\gamma$: however, the generator $S^{14}$ belongs to a doublet of $\SU(2)_L$, thus its presence will affect the breaking of the EW symmetry, and it will rotate in a non-trivial way the EW gauge generators.
From the above vacuum, the relation between the EW scale and the compositeness scale reads
\begin{equation}
\frac{v_{\rm SM}}{2 \sqrt{2} f} = 2 \cos \gamma\ \sin \frac{\theta}{2} \sqrt{1 - \cos^2 \gamma\ \sin^2 \frac{\theta}{2} } \equiv \sin (\tau)\,,
\end{equation}
where we define
 \begin{equation}
\sin \frac{\tau}{2} \equiv \cos \gamma\ \sin \frac{\theta}{2}\,.
\end{equation}
We thus see that the EW scale does depend on $\gamma$, and the angle $\tau$ replaces the function played by $\theta$ in describing the misalignment of the vacuum. We will thus systematically replace $\theta$ with $\tau$.

Once we redefine~\footnote{Note that $\cos \frac{\tau}{2} \equiv  \sqrt{\cos^2 \frac{\theta}{2} + \sin^2 \frac{\gamma}{2}\ \sin^2 \frac{\theta}{2}}=  \sqrt{\sin^2 \frac{\gamma}{2} + \cos^2 \frac{\gamma}{2}\ \cos^2 \frac{\theta}{2}}$.}
\begin{equation}
\begin{array}{c} 
h_1' = \frac{1}{\cos \frac{\tau}{2}} \left(\cos \gamma\ \cos \frac{\theta}{2}\ h_1 - \sin \gamma\ \varphi^0 \right)\,, \\
\varphi'_0 = \frac{1}{\cos \frac{\tau}{2}} \left(\sin \gamma\ h_1 + \cos \gamma\ \cos \frac{\theta}{2}\  \varphi_0 \right)\,,
\end{array}
\end{equation}
we find that only $h_1'$ couples to the $W$ and $Z$ bosons with couplings proportional to the SM ones
\begin{equation}
\frac{g_{h'_1 WW}}{g_{hWW}^{\rm SM} } = \frac{g_{h'_1 ZZ}}{g_{hZZ}^{\rm SM}} = \cos \tau\,,
\end{equation}
thus reproducing Eq.~\eqref{eq:ghWW}. In this basis, therefore, it is $h'_1$ that plays the role of the would-be Higgs boson.
Similarly, the Goldstones that give mass to $W$ and $Z$ are now a superposition of the $\pi_i$ and $\eta_3$ and $\eta^\pm$ (the explicit expressions can be found in Appendix~\ref{app:pNGB}). Thus, in the following the physical scalars will be denoted as $\eta'_3$ and ${\eta'}^\pm$.

In the SM fermion sector, using the new vacuum in the interaction in Eq.~\eqref{eq:topmass}, we can extract the mass of the top quark
\begin{equation} \label{topmass2}
m_t = 2 \cos \frac{\gamma}{2}\ \sin \frac{\theta}{2} \left( Y_{t2} \sin \frac{\gamma}{2} + Y_{t1} \cos \frac{\gamma}{2}\ \cos \frac{\theta}{2} \right) = f \sin (\tau) Y_{\rm top}\,,
\end{equation}
which leads to defining two new combinations of the couplings in Eq.~\eqref{eq:Yt1Yt2}:
\begin{equation} \label{eq:Ydef2}
Y_{\rm top} = \frac{1}{\cos \frac{\tau}{2}} \left( Y_{t2} \sin \gamma\ \sin \frac{\theta}{2} + Y_{t1}  \cos \frac{\theta}{2}\right)\,, \quad
Y_{\rm 0} = \frac{1}{\cos \frac{\tau}{2}} \left( Y_{t2}  \cos \frac{\theta}{2} - Y_{t1}\sin \gamma\ \sin \frac{\theta}{2}\right)\,.
\end{equation}
Note that the coupling $Y_{\rm top}$ is fixed by requiring the observed value of the top mass.
The Yukawa couplings will also include couplings of the tops to pNGBs, that include
\begin{equation}
- (t_L t_R^c) \left\{ \frac{Y_{\rm top}}{2 \sqrt{2}} c_\tau\ h'_1 + \frac{Y_0}{2 \sqrt{2}} (s_\frac{\tau}{2}\ \varphi'_0 + c_\frac{\tau}{2}\ h_2) + \dots  \right\} + \dots
\end{equation}
where we omitted couplings to other pNGBs and higher order ones coming from non-linearities. We see that the coupling of $h'_1$ differs from the ones of a SM Higgs by a factor $\cos \tau$, however, contrary to the case of gauge bosons, the other two scalars also couple to $t\bar{t}$.
This is relevant because, as we will see in the next section, the three scalars $h'_1$, $h_2$ and $\varphi'_0$ mix via mass terms.

%%%%%%%%%%%%%%%%%%%%
\subsection{Strategy for the minimisation of the potential} \label{sec:minimiz}

The potential that fixes the vacuum alignment is the same as in the previous section, with the difference of inserting the new $\gamma$-dependent vacuum: it thus includes the top loops, Eq.~\eqref{eq:Vtop}, the techni-fermion mass contribution, Eq.~\eqref{eq:Vm}, and the gauge loops, Eq.~\eqref{eq:Vgauge}. As discussed in the previous section, we will assume real Yukawas so that the tadpoles for the pseudo-scalars are absent.
The potential, expressed in terms of $\tau$ and up to constant terms, reads
\begin{align}
V (\tau, \beta, \gamma) = - C_t f^4 Y_t^2 s_\tau^2 - C_g f^4 \frac{3 g^2 + {g'}^2}{2} c_\tau^2
+  16 B \left( 2 M \frac{1- \Delta s_\gamma^2}{c_\gamma^2} + \delta m_R\ c_{2\beta} \right) s_\frac{\tau}{2}^2\,. 
\end{align}
We remark that the potential depends explicitly on $\gamma$ only for $\Delta \neq 1$ while $\beta$ always comes with $\delta m_R$, while an implicit dependence is hidden in the definition of $Y_t$. From the same expressions, we can compute the tadpoles for the three scalars $h'_1$, $h_2$ and $\varphi'_0$, and setting them to zero will determine the value of the three angles at the minimum.
In fact, we note that the tadpoles are always proportional to derivatives of the potential with respect to the angles. Explicitly, we find:
\begin{eqnarray}
\mbox{Tadpole}(h_2): & \Rightarrow& \frac{1}{2\sqrt{2} f} \frac{1}{2 s_\frac{\tau}{2}} \frac{\partial V}{\partial \beta}\,, \nonumber\\
\mbox{Tadpole}(h'_1): & \Rightarrow& \frac{1}{2\sqrt{2} f} \left( c_\gamma c_\frac{\theta}{2} \frac{\partial V}{\partial \theta} - s_\gamma  \frac{c_\gamma}{2 s_\frac{\tau}{2}} \frac{\partial V}{\partial \gamma} \right)\,, \\
\mbox{Tadpole}(\varphi'_0): & \Rightarrow& \frac{1}{2\sqrt{2} f} \left( s_\gamma  \frac{\partial V}{\partial \theta} + c_\gamma c_\frac{\theta}{2}  \frac{c_\gamma}{2 s_\frac{\tau}{2}} \frac{\partial V}{\partial \gamma} \right) \,. \nonumber 
\end{eqnarray}
Instead of determining the angles at the minimum for every choice of the input parameters, in order to simplify the analysis our strategy is to determine three of the initial parameters in the potential as a function of the angles. Then, we check the allowed range of values of the three angles for which the solution is indeed a minimum. This last requirement is equivalent to checking that no tachyons are present in the pNGB spectrum.
The most convenient choice of parameters is $Y_0$, $\Delta$ and $M$, for which we find the following relations:
\begin{eqnarray}
Y_0 &=& - \frac{2 K_\delta s_{2 \beta}}{Y_t c_\frac{\tau}{2}^2}\,, \label{eq:tadY0} \\
\Delta &=& 1 + \frac{C_t f^4}{4 B M} \frac{K_\delta s_{2\beta} s_\frac{\tau}{2} c_\gamma^2}{\sqrt{2(c_{2\gamma} + c_\tau)} s_\gamma}\,, \label{eq:tadDelta}\\
\frac{8 B M}{C_t f^4} &=& \left( 1 - \frac{C_g}{C_t} \frac{3 g^2 + {g'}^2}{2} \right) c_\tau - 2 K_\delta \left( c_{2\beta} - s_{2\beta} \frac{2 s_\gamma s_\frac{\tau}{2}^3}{\sqrt{2(c_{2\gamma} + c_\tau)} c_\frac{\tau}{2}^2} \right)\,.  \label{eq:tadM}
\end{eqnarray}
Besides the three angles, the only free parameters are $K_\delta$ and $C_g$, as $Y_t$ is fixed by the top mass and $C_t$ by fitting the Higgs mass.
We remark that for $\delta m_R=0$ (i.e., $K_\delta=0$), the dependence on $\beta$ disappear (besides the implicit dependence in $Y_t$): as we have seen in the previous section, in this case $\beta$ can be removed and the vacuum described in terms of a single angle $\tau \equiv \theta$.
Another special point is for $\Delta = 1$, which corresponds to $m_R = 0$ (i.e., $m_{R2} = - m_{R1}$). From Eq.(\ref{eq:tadDelta}) it follows that either $K_\delta=0$ or $\beta=0, \pi/2$ (and $Y_0=0$ determines the value of $\gamma$). However, we checked that there always is a tachyon in the pseudo-scalar spectrum, unless $K_\delta=0$, signalling that the wrong vacuum has been selected. For $K_\delta=0$, the vacuum can be aligned to the one already described in the previous section.
This simple analysis confirms that both $\beta$ and $\gamma$ are required to be non-zero.

One of the most interesting features of this vacuum is that a mass mixing between the three scalars $h'_1$, $h_2$ and $\varphi_0'$ is always generated by the potential. Due to the mixing with heavier states, the mass of the would-be Higgs $h'_1$ is parametrically reduced compared to the value in the $\theta$-vacuum. As a consequence, the value of the low energy constant $C_t$ that determines its value can be larger than what we found in the previous section, in Eq.(\ref{eq:mh0}), and it is not fixed to a single value any more.
This kind of mechanism was proposed to reduce the fine tuning in the Higgs mass in the minimal model $\SU(4)/\Sp(4)$ in Refs~\cite{Serra:2015xfa,Banerjee:2017qod}, where the authors added an additional parameter in the partial compositeness mixing of the right-handed top quark. In the $\SU(6)/\Sp(6)$ case we consider, however, the mixing arises naturally from the potential. Furthermore, the mixing does not involve CP-odd scalars~\cite{Alanne:2018wtp}. 
The complete mass matrix for the three scalars is reported in Appendix~\ref{app:pNGB}, and it contains mixing proportional to $K_\delta$ between the would-be Higgs $h'_1$ and the other two: remarkably, the one with the EW singlet $\varphi'_0$ arises at order $s_\tau$ in an expansion for small $\tau$.
It is instructive to expand the mass spectrum for small $s_\tau$ and small $K_\delta$. Besides one state (approximately associated to $h_2$) that receives a mass of order $f$, i.e. unsuppressed by $s_\tau$, the spectrum contains two lighter states with masses:
\begin{align}
m_{1/2}^2 = \frac{C_t f^2 s_\tau^2}{8} \left[ \left(Y_t^2 - \frac{C_g}{C_t} \frac{3 g^2 + {g'}^2}{2} \right) Q_\pm (\gamma) - K_\delta c_{2\beta}\ Q_\mp (\gamma) + \mathcal{O} (K_\delta^2) \right] \nonumber \\
\mp \frac{C_t f^2 s_\tau}{8} K_\delta s_{2\beta} \sqrt{\cot^2 \gamma + 16}\ Q_\mp (\gamma) + \mathcal{O} (s_\tau)^3\,,
\end{align}
where the functions $Q_\pm (\gamma)$ are defined as
\begin{equation}
Q_{\pm} (\gamma) = 1 \pm \frac{\cot{\gamma}}{\sqrt{\cot^2 \gamma + 16}}\,.
\end{equation}
For $\gamma \to 0^+$, we note that $Q_+ \to 2$ while $Q_- \to 0 + \mathcal{O} (\gamma^2)$, while the roles are exchanged for $\gamma \to 0^-$. Thus, the would be Higgs can be identified with $m_1$ for positive $\gamma$, and $m_2$ for negative $\gamma$. 
For $\gamma \sim 0$, therefore, the mass of the would-be Higgs recovers the result we obtained in the previous section.
For non-zero $\gamma$, however, an additional correction of order $s_\tau$ is induced by the term in the second line.
This can be either positive or negative, depending on the sign of $\beta$ and $\delta m_R$, thus it can allow a larger or smaller value of $C_t$ compared to $C_t\sim 2$. We recall here that $C_t$ is determined by the strong dynamics and by the interactions in the UV that generate the top mass operator, so that it helps to have some flexibility regarding its value.
The most interesting point is that the additional correction is enhanced at small $\tau$, thus it is more important for small alignment angles.

In the following section, we will study this case numerically to determine the allowed range for the coefficient $C_t$ that fits the Higgs mass. In fact, constraints arise both from the range of validity of the parameters, and from the modifications of the would-be Higgs couplings due to the mixing between $h'_1$ and $\varphi'_0$.

%%%%%%%%%%%%%%%%%%%%%%%%%%%%%%%%%%%%%%%
\subsection{Numerical results}

\begin{figure}[tb!]
\begin{center}
\includegraphics[width=16cm]{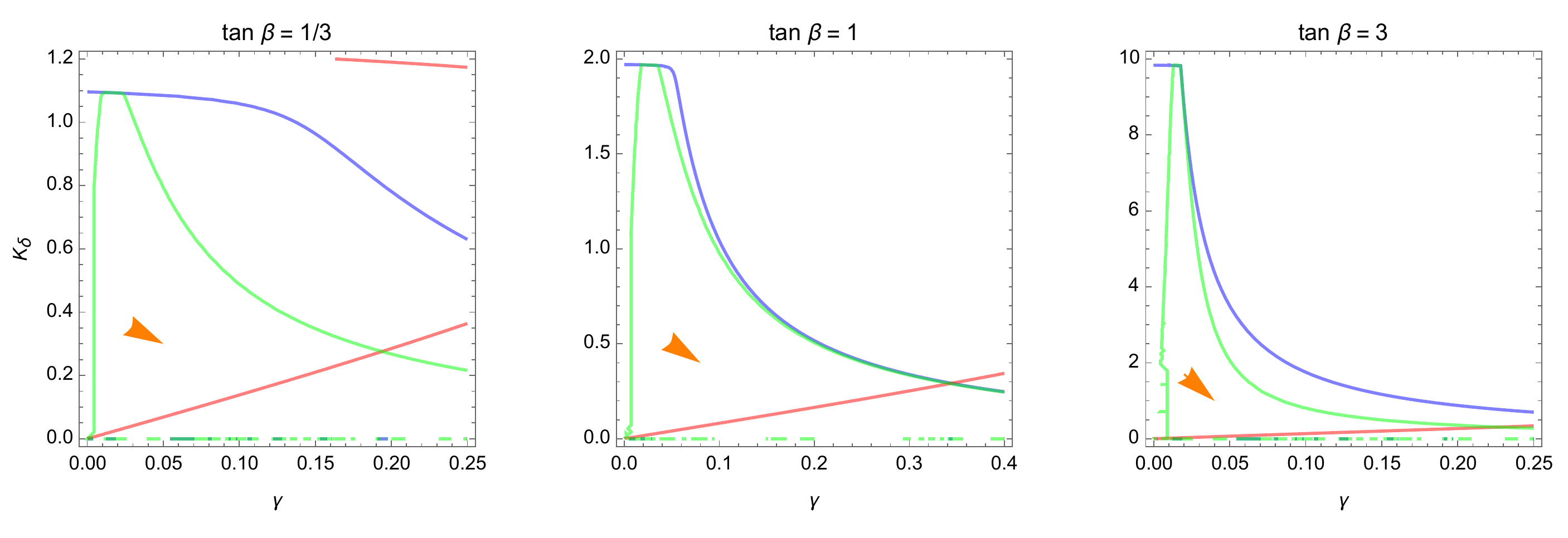} 
\end{center}
\caption{The allowed parameter space for $\tau=0.1$ and $C_g = 1/3\ C_t$ is indicated by the orange arrowhead. The blue, green and red contours indicate where the neutral scalars, pseudo-scalars and charged scalars respectively develop a tachyonic spectrum. Below the red line, the theory will fall to a charge-violating vacuum.} \label{fig:paramSpace}
\end{figure}

For concreteness, we focus on a benchmark model with $\tau = 0.1$, corresponding to a compositeness scale of $2 \sqrt{2} f = 2.4$~TeV, and fix $C_g = 1/3\ C_t$. We then study the parameter space as a function of $\gamma$, $\beta$ and $K_\delta$: the allowed one is given by the region where no tachyons exist in the spectrum.
The result is shown in Fig.~\ref{fig:paramSpace}, where the three colours delimitate the safe regions for the neutral scalars (blue), pseudo-scalars (green) and charged ones (red). We see that the allowed region, indicated by an orange arrowhead, is mainly determined by the pseudo-scalars and charged scalars developing tachyons. Furthermore, the region with $\gamma \to 0$ is not allowed due to pseudo-scalars, while $K_\delta \to 0$ is prevented by the charged ones. At the origin, $\gamma=0$ and $K_\delta = 0$, the vacuum collapses to the one described in the previous section, where $\beta$ can be rotated away. We remark that the red line delimits the region where charged tachyons appear: this implies that for parameter values below the red line the vacuum will be misaligned along a charge-violating direction.
\begin{figure}[tb!]
\begin{center}
\includegraphics[width=16cm]{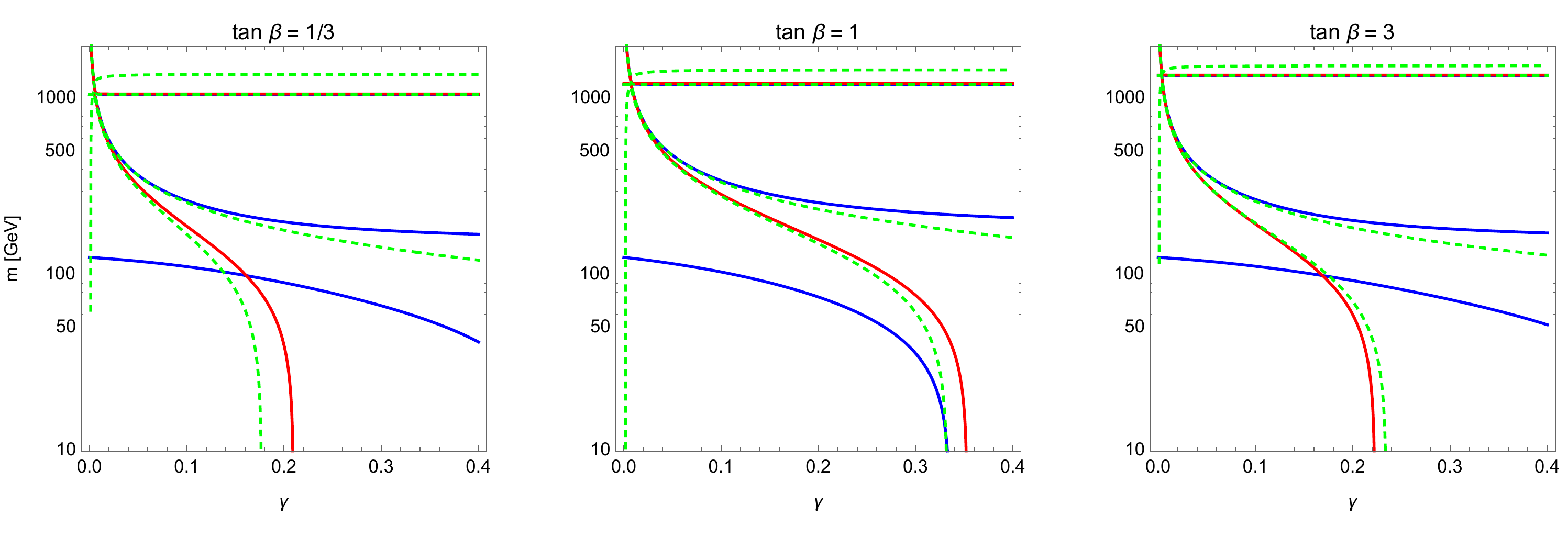} 
\end{center}
\caption{Spectra for $\tau=0.1$, $C_t=2$ and $C_g = 1/3\ C_t$. The blue, green and red lines correspond to neutral scalars, pseudo-scalars and charged scalars respectively. We fix $K_\delta = 0.3$ and show spectra for $\tan \beta = 1/3$, $1$ and $3$.} \label{fig:Spectra1}
\end{figure}

To study the spectrum, we find useful to slice the parameter space further and show characteristic plots as a function of a single variable. In Fig.~\ref{fig:Spectra1} we fix $K_\delta = 0.3$ and show the masses of the neutral scalars (blue), pseudo-scalars (green) and charged scalars (red) as a function of $\gamma$ for $\tan \beta = 1/3$, $1$ and $3$. We fix $C_t=2$ so that for small $\gamma$ the Higgs mass is recovered for the lightest neutral scalar. We recognise that very small values of $\gamma$ are ruled out by a pseudo-scalar becoming tachyonic, while the maximal allowed value of $\gamma$ depends on a pseudo-scalar becoming tachyonic for $\tan \beta < 1$, and a charged scalar for $\tan \beta > 1$. Note also that the heaviest neutral scalar is nearly degenerate with the heavier charged one and with one of the pseudo-scalars.
We also remark significant reduction of the would-be Higgs mass for non-zero values of $\gamma$: this needs to raise the value of $C_t \propto 1/m^2$ in order to keep the fit with the experimentally measured mass. This plot already shows a proof of principle that the mixing can allow for coefficients larger than 2. We also note that $\tan \beta \sim 1$ is favoured, because it allows a larger variation of the value of the mass. This is possible because, for the choice of $K_\delta$ we made, the line where the pseudo-scalar spectrum becomes tachyonic is close to the line where the lightest scalar mass approaches zero, as it can be seen in the central plot of Fig.~\ref{fig:paramSpace}.
\begin{figure}[tb!]
\begin{center}
\includegraphics[width=16cm]{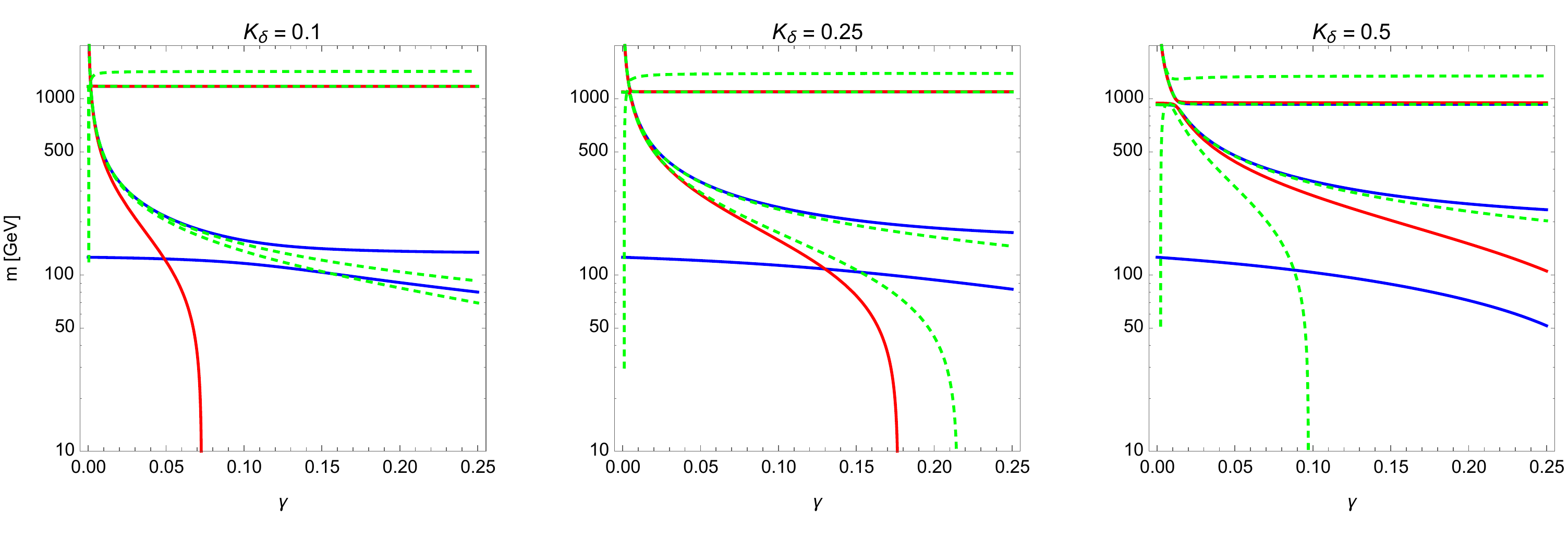} 
\end{center}
\caption{Same as Fig.~\ref{fig:Spectra1} for fixed $\tan \beta = 1/3$, showing spectra for $K_\delta = 0.1$, $0.25$ and $0.5$.} \label{fig:Spectra2}
\end{figure}
\begin{figure}[tb!]
\begin{center}
\includegraphics[width=16cm]{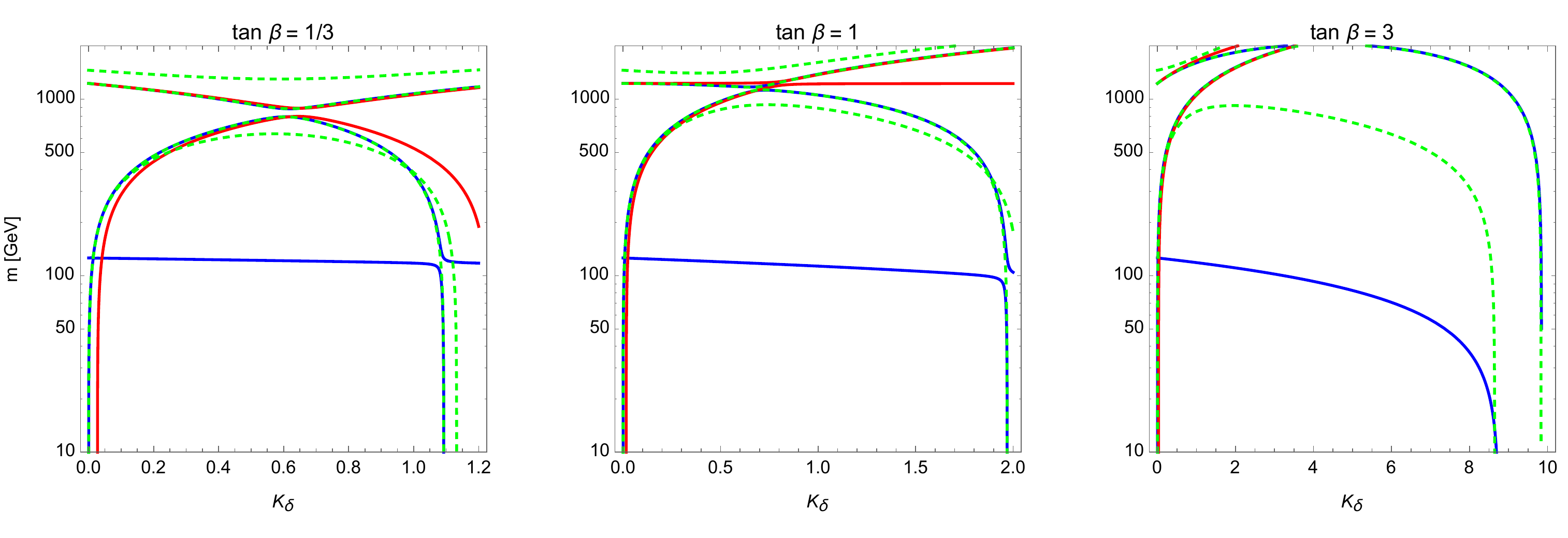} 
\end{center}
\caption{Same as Figs~\ref{fig:Spectra1} and~~\ref{fig:Spectra2} for fixed $\gamma = 0.02$, showing spectra for $\tan \beta = 1/3$, $1$ and $3$.} \label{fig:Spectra3}
\end{figure}

For completeness, in Figs~\ref{fig:Spectra2} and~\ref{fig:Spectra3}, we show the spectra as a function of $\gamma$ for $\tan \beta = 1/3$ and $K_\delta = 0.1$, $0.25$ and $0.5$, and as a function of $K_\delta$ for $\gamma=0.02$ and $\tan \beta = 1/3$, $1$ and $3$. A common feature of all these plots is the presence of light scalars, whose masses are below $1$~TeV, even if the condensation scale of the benchmark point we chose is $2.4$~TeV. Thus, the presence of these additional scalars may well be accessible at the LHC. The non-vanishing couplings to tops, in fact, will guarantee potentially observable production rates via loop-induced gluon fusion for the neutral states.

%%%%%%%%%%%%%%%%%%%%%%%%%%%%%%%%%%%%%%
\section{Final remarks and conclusions} \label{sec:concl}

In this paper, we have analysed in detail the vacuum alignment for a model of composite Higgs based on the coset $\SU(6)/\Sp(6)$. In terms of the underlying theory, this is a simple generalisation of the minimal case $\SU(4)/\Sp(4)$ by the addition of one further Dirac fermion. However, the global symmetry is significantly enlarged, allowing for a second Higgs doublet in the pNGB spectrum.  We focus in particular on a model with a single $\SU(2)_L$ doublet in order to avoid the presence of EW triplets in the spectrum. The underlying fermions thus consist of one $SU(2)_L$ doublet and two $\SU(2)_R$ doublets.

We studied the vacuum alignment problem in presence of bilinear operators at the origin of the SM fermion masses, but we checked that our conclusions also apply to models with partial compositeness. In absence of CP violating phases, we find that the vacuum can be misaligned only in two exclusive ways, depending on the  alignment of the Yukawa couplings and underlying fermion masses:

\begin{itemize}

\item[a)] The misalignment is characterised by a single angle $\theta$, which corresponds to a rotation along one of the two Higgs doublets. This situation is achieved if the Yukawas are aligned in the $\SU(6)$ space and the two $\SU(2)_R$ doublets are degenerate, or if the Yukawas only involve one of the two $\SU(2)_R$ doublets.

\item[b)] The misalignment depends on 3 angles, $\theta$, $\beta$ and $\gamma$. While the first two correspond to an ordinary two-Higgs-doublet model, $\gamma$ is generated by a singlet. All three angles must be non-vanishing in this scenario.

\end{itemize}

In the first case a), a global $\U(1)$ symmetry remains unbroken, thus a subset of pNGBs is prevented from decaying into a pair of SM states. The spectrum therefore contains two sets of nearly degenerate neutral and electromagnetically charged states that carry the global $\U(1)$ charge, and the lightest one can be associated with a Dark Matter candidate. Besides the would-be Higgs, which has properties similar to the minimal $\SU(4)/\Sp(4)$ model, the spectrum also contains two pseudo-scalars that decay into SM gauge bosons via topological anomalies. We provide details of the spectrum of the theory, where a pseudo-scalar lighter than the DM candidate is a common occurrence, and a complete list of the relevant couplings among pNGBs. The detailed study of the properties of the Dark Matter candidate are left for future investigations, with all the necessary tools already provided in this paper.

In the second case, there is no stable pNGB, and all the neutral scalars, pseudo-scalars and charged scalars mix with each other. In particular, a mixing between the would-be Higgs boson and a scalar singlet is always generated, and its presence tends to parametrically reduce the value of the physical Higgs mass at the price of facing constraints from the consequent reduction of the Higgs couplings to SM states. Note that the current determination of the Higgs couplings at the LHC still allows for a $10\div 20$\% deviation. This mechanism, already discussed in the literature, has the potential of reducing the tuning in the Higgs mass by allowing for larger form factors coming from the strong dynamics. The novelty of this result in the $\SU(6)/\Sp(6)$ model is that this mixing is not introduced ad-hoc, but it is a necessary feature of the most general vacuum. Furthermore, the mixing only involves CP-even scalars, contrary to the case in the minimal $\SU(4)/\Sp(4)$ model that has been studied in the literature.
Once lattice results determining the value of the form factors are available, our results will allow to further constrain the parameter space of the model and thus predict the phenomenology of the spectrum of scalars. One common feature we already identify is the presence of light states, with masses below the TeV, even when the compositeness scale is pushed in the multi-TeV range.

%%%%%%%%%%%%%%%%%%%%%%%%%%%%%%%%%%%%%%
\section*{Acknowledgements}

We thank M.Lespinasse for his contribution in the initial stages of this project.
GC acknowledges partial support from the Labex-LIO (Lyon Institute of Origins) under grant ANR-10-LABX-66 (Agence Nationale de la Recherche) and FRAMA (FR3127, F\'ed\'eration de Recherche ``Andr\'e Marie Amp\`ere''). This work has also been supported by the LIA FCPPL (France-China Particle Physics Laboratory). 
HHZ and CC are supported by the National Natural Science Foundation of China (NSFC) under Grant Nos. 11875327 and 11375277, the China Postdoctoral Science Foundation under Grant No. 2018M643282, the Natural Science Foundation of Guangdong Province under Grant no. 2016A030313313, and the Sun Yat-Sen University Science Foundation.
GC thanks the Sun Yat-Sen University for hospitality during the completion of this work.

\appendix

\section{Properties of the model}

\subsection{Broken and un-broken generators in the vacuum $\Sigma_0$} \label{app:generators}

The 21 unbroken generators of Sp(6) are:
{\tiny
\begin{equation}
S^1 = \frac{1}{2}\begin{pmatrix}
\sigma_1 & 0 & 0\\
0 & 0 & 0\\
0 & 0 & 0
\end{pmatrix}\,,
\quad 
S^2 = \frac{1}{2}\begin{pmatrix}
\sigma_2 & 0 & 0\\
0 & 0 & 0\\
0 & 0 & 0
\end{pmatrix}\,,
\quad 
S^3 = \frac{1}{2}\begin{pmatrix}
\sigma_3 & 0 & 0\\
0 & 0 & 0\\
0 & 0 & 0
\end{pmatrix}\,,
\end{equation}
\begin{equation}
S^4 = \frac{1}{2}\begin{pmatrix}
0 & 0 & 0\\
0 & -\sigma_1^T & 0\\
0 & 0 & 0
\end{pmatrix}\,,
\quad 
S^5 = \frac{1}{2}\begin{pmatrix}
0 & 0 & 0\\
0 & -\sigma_2^T & 0\\
0 & 0 & 0
\end{pmatrix}\,,
\quad 
S^6 = \frac{1}{2}\begin{pmatrix}
0 & 0 & 0\\
0 & -\sigma_3^T & 0\\
0 & 0 & 0
\end{pmatrix}\,,
\end{equation}
\begin{equation}
S^7 = \frac{1}{2}\begin{pmatrix}
0 & 0 & 0\\
0 & 0 & 0\\
0 & 0 & -\sigma_1^T
\end{pmatrix}\,,
\quad 
S^8 = \frac{1}{2}\begin{pmatrix}
0 & 0 & 0\\
0 & 0 & 0\\
0 & 0 & -\sigma_2^T
\end{pmatrix}\,,
\quad 
S^9 = \frac{1}{2}\begin{pmatrix}
0 & 0 & 0\\
0 & 0 & 0\\
0 & 0 & -\sigma_3^T
\end{pmatrix}\,,
\end{equation}
\begin{equation}
S^{10} = \frac{1}{2\sqrt{2}}\begin{pmatrix}
0 & i \sigma_1 & 0\\
- i \sigma_1 & 0 & 0\\
0 & 0 & 0
\end{pmatrix}\,,
\quad 
S^{11} = \frac{1}{2\sqrt{2}}\begin{pmatrix}
0 & i \sigma_2 & 0\\
-i \sigma_2 & 0 & 0\\
0 & 0 & 0
\end{pmatrix}\,
\quad 
S^{12} = \frac{1}{2\sqrt{2}}\begin{pmatrix}
0 & i \sigma_3 & 0\\
-i \sigma_3 & 0 & 0\\
0 & 0 & 0
\end{pmatrix}\,,
\quad 
S^{13} = \frac{1}{2\sqrt{2}}\begin{pmatrix}
0 & \mathbb{I}_2 & 0\\
\mathbb{I}_2 & 0 & 0\\
0 & 0 & 0
\end{pmatrix}\,,
\end{equation}
\begin{equation}
S^{14} = \frac{1}{2\sqrt{2}}\begin{pmatrix}
0 & 0 & i \sigma_1\\
0 & 0 & 0\\
- i \sigma_1 & 0 & 0
\end{pmatrix}\,,
\quad 
S^{15} = \frac{1}{2\sqrt{2}}\begin{pmatrix}
0 & 0 & i \sigma_2 \\
0 & 0 & 0\\
-i \sigma_2 & 0 & 0
\end{pmatrix}\,,
\quad 
S^{16} = \frac{1}{2\sqrt{2}}\begin{pmatrix}
0 & 0 & i \sigma_3\\
0 & 0 & 0\\
-i \sigma_3 & 0 & 0
\end{pmatrix}\,,
\quad 
S^{17} = \frac{1}{2\sqrt{2}}\begin{pmatrix}
0 & 0 & \mathbb{I}_2\\
0 & 0 & 0\\
\mathbb{I}_2 & 0 & 0
\end{pmatrix}\,,
\end{equation}
\begin{equation}
S^{18} = \frac{1}{2\sqrt{2}}\begin{pmatrix}
0 & 0 & 0\\
0 & 0 & \sigma_1\\
0 & \sigma_1 & 0
\end{pmatrix}\,,
\quad 
S^{19} = \frac{1}{2\sqrt{2}}\begin{pmatrix}
0 & 0 & 0 \\
0 & 0 & \sigma_2\\
0 & \sigma_2 & 0
\end{pmatrix}\,,
\quad 
S^{20} = \frac{1}{2\sqrt{2}}\begin{pmatrix}
0 & 0 & 0\\
0 & 0 & \sigma_3\\
0 & \sigma_3 & 0
\end{pmatrix}\,,
\quad 
S^{21} = \frac{1}{2\sqrt{2}}\begin{pmatrix}
0 & 0 & 0\\
0 & 0 & i \mathbb{I}_2\\
0 & -i \mathbb{I}_2 & 0
\end{pmatrix}\,,
\end{equation}}
The first 3 correspond to the gauged SU(2)$_L$, while the following 6 to the two partly-gauged SU(2)$_R$'s.
The 3 sets of generators $S^{10,\dots 13}$, $S^{14,\dots 17}$ and $S^{18,\dots 21}$ transform as bidoublets.

The 14 broken generators are:
{\tiny
\begin{equation}
X^1 = \frac{1}{2\sqrt{2}}\begin{pmatrix}
\mathbb{I}_2 & 0 & 0\\
0 & -\mathbb{I}_2 & 0\\
0 & 0 & 0
\end{pmatrix}\,,
\quad 
X^2 = \frac{1}{2\sqrt{6}}\begin{pmatrix}
\mathbb{I}_2 & 0 & 0\\
0 & \mathbb{I}_2 & 0\\
0 & 0 & -2 \mathbb{I}_2
\end{pmatrix}\,,
\end{equation}
\begin{equation}
X^{3} = \frac{1}{2\sqrt{2}}\begin{pmatrix}
0 &  \sigma_1 & 0\\
 \sigma_1 & 0 & 0\\
0 & 0 & 0
\end{pmatrix}\,,
\quad 
X^{4} = \frac{1}{2\sqrt{2}}\begin{pmatrix}
0 &  \sigma_2 & 0\\
 \sigma_2 & 0 & 0\\
0 & 0 & 0
\end{pmatrix}\,,
\quad 
X^{5} = \frac{1}{2\sqrt{2}}\begin{pmatrix}
0 &  \sigma_3 & 0\\
 \sigma_3 & 0 & 0\\
0 & 0 & 0
\end{pmatrix}\,,
\quad 
X^{6} = \frac{1}{2\sqrt{2}}\begin{pmatrix}
0 & i \mathbb{I}_2 & 0\\
- i \mathbb{I}_2 & 0 & 0\\
0 & 0 & 0
\end{pmatrix}\,,
\end{equation}
\begin{equation}
X^{7} = \frac{1}{2\sqrt{2}}\begin{pmatrix}
0 & 0 &  \sigma_1\\
0 & 0 & 0\\
 \sigma_1 & 0 & 0
\end{pmatrix}\,,
\quad 
X^{8} = \frac{1}{2\sqrt{2}}\begin{pmatrix}
0 & 0 &  \sigma_2 \\
0 & 0 & 0\\
 \sigma_2 & 0 & 0
\end{pmatrix}\,,
\quad 
X^{9} = \frac{1}{2\sqrt{2}}\begin{pmatrix}
0 & 0 &  \sigma_3\\
0 & 0 & 0\\
 \sigma_3 & 0 & 0
\end{pmatrix}\,,
\quad 
X^{10} = \frac{1}{2\sqrt{2}}\begin{pmatrix}
0 & 0 & i \mathbb{I}_2\\
0 & 0 & 0\\
-i \mathbb{I}_2 & 0 & 0
\end{pmatrix}\,,
\end{equation}
\begin{equation}
X^{11} = \frac{1}{2\sqrt{2}}\begin{pmatrix}
0 & 0 & 0\\
0 & 0 & i \sigma_1\\
0 & -i \sigma_1 & 0
\end{pmatrix}\,,
\quad 
X^{12} = \frac{1}{2\sqrt{2}}\begin{pmatrix}
0 & 0 & 0 \\
0 & 0 & i\sigma_2\\
0 & -i\sigma_2 & 0
\end{pmatrix}\,,
\quad 
X^{13} = \frac{1}{2\sqrt{2}}\begin{pmatrix}
0 & 0 & 0\\
0 & 0 & i\sigma_3\\
0 & -i\sigma_3 & 0
\end{pmatrix}\,,
\quad 
X^{14} = \frac{1}{2\sqrt{2}}\begin{pmatrix}
0 & 0 & 0\\
0 & 0 & \mathbb{I}_2\\
0 & \mathbb{I}_2 & 0
\end{pmatrix}\,.
\end{equation}}
The first two correspond to the singlets, while the following 3 groups of 4 (in each row) correspond to the 3 bi-doublets.

\subsection{CP properties and Wess-Zumino-Witten topological term} \label{app:CP}

The Wess-Zumino-Witten topological term~\cite{Wess:1971yu,Witten:1983tw} reads:
\begin{eqnarray}
\mathcal{L}_{WZW}=\frac{d_{\rm FCD} g_{V_1V_2}}{16\sqrt{2}\pi^2f} \left(c_\theta\ \eta_1 + \frac{1}{\sqrt{3} c_\theta}\ \eta_2 \right)\ \epsilon_{\mu\nu\rho\sigma}V_1^{\mu\nu}V_2^{\rho\sigma}\,,
\end{eqnarray}
where $d_{\rm FCD}$ is the dimension of the FCD representation of the underlying fermions ($d_{\rm FCD} = 2$ in the minimal $\SU(2)_{\rm TC}$ model), and
\begin{equation}
g_{WW}=g^2\,, \quad g_{ZZ}=(g^2-{g'}^2),\quad g_{Z\gamma}=g g'\,.
\end{equation}
This result shows that the model has the same anomaly structure as the minimal $\SU(4)/\Sp(4)$ model~\cite{Arbey:2015exa}, as a linear combination of the two singlets have the same WZW couplings. In particular, note the absence of couplings to two photons.
The couplings above also show that $\eta_1$ and $\eta_2$ are pseudo-scalars under CP.

\subsection{Lowest order pNGB couplings} \label{app:pNGBcoupl}

The chiral Lagrangian at LO contains couplings to one and two gauge bosons, via the covariant derivatives. In the following, we use the short notation $c_W=\cos\theta_W,\ c_{2W}=\cos2\theta_W, \ t_W=\tan\theta_W$ and drop the exact Goldstones eaten by the $W$ and $Z$ (in the $\theta$--$\beta$ vacuum).
Also, for short notation, we define (Cf. Section~\ref{sec:i2HDM}):
\begin{equation}
H^0 = \frac{h_2 - i A_0}{\sqrt{2}}\,, \; \eta^0 = \frac{\varphi_0 - i \eta_3}{\sqrt{2}}\,.
\end{equation}

The $V\partial\phi \phi$ interactions read
\begin{eqnarray}
\mathcal{L}_{V\partial\phi\phi}&=&\frac{g}{2\sqrt{2}}W_\mu^+\ [(1+c_\theta)\ H^-i\overset\leftrightarrow{\partial^\mu} H^0 - (1-c_\theta)\ \eta^-i\overset\leftrightarrow{\partial^\mu}\eta^0]+\mbox{h.c.}\nonumber\\
&&+\frac{g}{4c_W}Z_\mu\ [(1+c_\theta-4s_W^2)\ H^-i\overset\leftrightarrow{\partial^\mu} H^+ + (1-c_\theta-4s_W^2)\ \eta^-i\overset\leftrightarrow{\partial^\mu} \eta^+\nonumber\\
&&\qquad\qquad+(1+c_\theta)\ H^0i\overset\leftrightarrow{\partial^\mu} (H^0)^\ast + (1-c_\theta)\ \eta^0i\overset\leftrightarrow{\partial^\mu} (\eta^0)^\ast]\nonumber\\
&&+eA_\mu\ [H^-i\overset\leftrightarrow{\partial^\mu} H^+ + \eta^-i\overset\leftrightarrow{\partial^\mu} \eta^+]\,;
\end{eqnarray}
where $\phi_1 \overset\leftrightarrow{\partial^\mu} \phi_2 = \phi_1 (\partial^\mu \phi_2) - (\partial^\mu \phi_1) \phi_2$.

The $VV\phi\phi$ interactions read
\begin{eqnarray}
\mathcal{L}_{VV\phi\phi}&=&e^2A^\mu A_\mu\ [H^+H^-+\eta^+\eta^-]\nonumber\\
&&+\frac{g^2}{4}W^+_\mu W^{-,\mu}\ [c_{2\theta} h_1^2-s_\theta^2\eta_1^2+(1+c_\theta)c_\theta(|H^0|^2+H^+H^-) - (1-c_\theta)c_\theta(|\eta^0|^2+\eta^+\eta^-)]\nonumber\\
&&+\frac{g^2}{8c_W^2}Z^\mu Z_\mu\ [(2c_{2W}^2-(1-c_\theta)(c_\theta+2c_{2W}))H^+H^- +(2c_{2W}^2+(1+c_\theta)(c_\theta-2c_{2W}))\eta^+\eta^-\nonumber\\
&&\phantom{+\frac{g^2}{8c_W^2}Z_\mu Z^\mu} +(1+c_\theta)c_\theta|H^0|^2-(1-c_\theta)c_\theta|\eta^0|^2+c_{2\theta}h_1^2-s_\theta^2\eta_1^2]\nonumber\\
&&+\frac{eg}{2c_W}A^\mu Z_\mu\ [(2c_{2W}-1+c_\theta)H^+H^-+(2c_{2W}-1-c_\theta)\eta^+\eta^-]\nonumber\\
&&+\frac{eg}{2\sqrt{2}}A^\mu W^+_\mu\ [(1+c_\theta)H^-H^0-(1-c_\theta)\eta^-\eta^0]+\mbox{h.c.}\nonumber\\
&&-\frac{g^2}{2\sqrt{2}c_W}Z^\mu W^+_\mu\  [(1+c_\theta)H^-H^0+(1-c_\theta)\eta^-\eta^0]+ \mbox{h.c.}
\end{eqnarray}
In addition, there are couplings of two (massive) vectors with the Higgs boson
\begin{eqnarray}
\mathcal{L}_{VVh_1}=\sqrt{2}g^2fs_\theta c_\theta\ W_\mu^+W^{-,\mu}\ h_1+\frac{(g'^2+g^2)f}{\sqrt{2}}s_\theta c_\theta\ Z_\mu Z^\mu\ h_1\,.
\end{eqnarray}

The chiral Lagrangian also contains self-interactions of the pNGBs, starting with quartic terms. Here we will only report the couplings containing the Higgs boson, as they are the most relevant ones for the production at collider of the additional scalars, and for studying the properties of the Dark Matter candidate.
The relevant interactions consist of couplings bilinear in the Higgs boson $h_1$, that read
\begin{equation}
\mathcal{L}_{h1^2 \phi \phi \partial^2} = \frac{1}{96 f^2} \left[ h_1 \partial_\mu h_1\ (\phi^\ast \partial^\mu \phi + \phi \partial^\mu \phi^\ast) - h_1^2\ \partial_\mu \phi^\ast \partial^\mu \phi - \partial_\mu h_1 \partial^\mu h_1\ \phi^\ast \phi \right]\,;
\end{equation}
where $\phi = H^+, \eta^+, H^0, \eta^0, \sqrt{2} \eta_1$. The couplings linear in the Higgs read
\begin{eqnarray}
\mathcal{L}_{h_1 \phi^3 \partial^2} &=& -\frac{i}{32f^2}(h_1\overset\leftrightarrow{\partial^\mu}\eta_1)\ [\eta^-\overset\leftrightarrow{\partial_\mu} H^+ + \eta^0\overset\leftrightarrow{\partial_\mu} (H^0)^\ast] + \mbox{h.c.} \nonumber\\
&& -\frac{i\sqrt{3}}{96f^2}(h_1\partial^\mu\eta_2+\eta_2\partial^\mu h_1)\ [H^-\partial_\mu \eta^+ + \eta^+\partial^\mu H^- + H^0\partial_\mu (\eta^0)^\ast + (\eta^0)^\ast\partial^\mu H^0] + \mbox{h.c.}\nonumber\\
&&+\frac{i}{48f^2}h_1\eta_2\ [\partial^\mu\eta^+\partial_\mu H^- + \partial^\mu(\eta^0)^\ast\partial_\mu H^0] + \mbox{h.c.}
\end{eqnarray}

%%%%%%%%%%%%%%%%%%%%%%%%%%%%%%%
\section{Spurions and the $\theta$--vacuum}

\subsection{Projectors for the top mass} \label{app:topproj}

The projectors used in Eq.~\ref{eq:topmass} to define the effective operator generating the top mass are:
{\footnotesize 
\begin{align}
P^1_1 & = \frac{1}{2}
\begin{pmatrix}
0&0&1&0&0&0\\
0&0&0&0&0&0\\
-1&0&0&0&0&0\\
0&0&0&0&0&0\\
0&0&0&0&0&0\\
0&0&0&0&0&0
\end{pmatrix}
\qquad \qquad 
P^2_1 = \frac{1}{2}
\begin{pmatrix}
0&0&0&0&0&0\\
0&0&1&0&0&0\\
0&-1&0&0&0&0\\
0&0&0&0&0&0\\
0&0&0&0&0&0\\
0&0&0&0&0&0
\end{pmatrix}\\
P^1_2 & = \frac{1}{2}
\begin{pmatrix}
0&0&0&0&1&0\\
0&0&0&0&0&0\\
0&0&0&0&0&0\\
0&0&0&0&0&0\\
-1&0&0&0&0&0\\
0&0&0&0&0&0
\end{pmatrix}
\qquad \qquad 
P^2_2 = \frac{1}{2}
\begin{pmatrix}
0&0&0&0&0&0\\
0&0&0&0&1&0\\
0&0&0&0&0&0\\
0&0&0&0&0&0\\
0&-1&0&0&0&0\\
0&0&0&0&0&0
\end{pmatrix}
\end{align}}
Similarly, for the bottom mass in Eq.~\ref{eq:botmass}, we have:
{\footnotesize
\begin{align}
P^1_{b1} & = \frac{1}{2}
\begin{pmatrix}
0&0&0&1&0&0\\
0&0&0&0&0&0\\
0&0&0&0&0&0\\
-1&0&0&0&0&0\\
0&0&0&0&0&0\\
0&0&0&0&0&0
\end{pmatrix}
\qquad \qquad 
P^2_{b1} = \frac{1}{2}
\begin{pmatrix}
0&0&0&0&0&0\\
0&0&0&1&0&0\\
0&0&0&0&0&0\\
0&-1&0&0&0&0\\
0&0&0&0&0&0\\
0&0&0&0&0&0
\end{pmatrix}\\
P^1_{b2} & = \frac{1}{2}
\begin{pmatrix}
0&0&0&0&0&1\\
0&0&0&0&0&0\\
0&0&0&0&0&0\\
0&0&0&0&0&0\\
0&0&0&0&0&0\\
-1&0&0&0&0&0
\end{pmatrix}
\qquad \qquad 
P^2_{b2} = \frac{1}{2}
\begin{pmatrix}
0&0&0&0&0&0\\
0&0&0&0&0&1\\
0&0&0&0&0&0\\
0&0&0&0&0&0\\
0&0&0&0&0&0\\
0&-1&0&0&0&0
\end{pmatrix}
\end{align}}

For the case of partial compositeness with the ${\bf 21}_{\SU(6)}$, the spurions containing the pre-Yukawas are give by the following matrices:
{\footnotesize
\begin{align}
S^1_L  =
\begin{pmatrix}
0&0&y_{L1}&0&y_{L2}&0\\
0&0&0&0&0&0\\
y_{L1}&0&0&0&0&0\\
0&0&0&0&0&0\\
y_{L2}&0&0&0&0&0\\
0&0&0&0&0&0
\end{pmatrix}
\qquad 
S^2_L = 
\begin{pmatrix}
0&0&0&0&0&0\\
0&0&y_{L1}&0&y_{L2}&0\\
0&y_{L1}&0&0&0&0\\
0&0&0&0&0&0\\
0&y_{L2}&0&0&0&0\\
0&0&0&0&0&0
\end{pmatrix} \qquad
S_R  = y_{R}
\begin{pmatrix}
0&0&0&0&0&0\\
0&0&0&0&0&0\\
0&0&0&1&0&0\\
0&0&1&0&0&0\\
0&0&0&0&0&1\\
0&0&0&0&1&0
\end{pmatrix}\,.
\end{align}}

\subsection{Couplings of the pNGBs to tops} \label{app:pNGBcoupl2}

Besides the linear couplings of the pNGBs to top and bottom, reported in Eqs~(\ref{eq:topmass}) and (\ref{eq:botmass}), the effective interactions contain higher order terms that become relevant for production and annihilation of the pNGBs. Here we report the couplings involving two pNGBs, which are relevant for Dark Matter and collider phenomenology:
\begin{eqnarray}
\mathcal{L}_{\phi\phi ff}&\supset&\frac{Y_{t1}}{16f}(t_Lt_R^c)^\dag\left[-\frac{2i}{\sqrt{3}}c_\theta h_1\eta_2+(1-c_\theta)(H^-\eta^++(\eta^0)^\ast H^0) + (1+c_\theta)(H^+\eta^-+\eta^0(H^0)^\ast)\right. \nonumber\\
&&\left.\phantom{+\frac{Y_{t1}}{f}(t_Lt_R^c)^\dag}+s_\theta\left(\eta^+\eta^-+H^+H^-+|H^0|^2+|\eta^0|^2+h_1^2+\eta_1^2+\frac{1}{3}\eta_2^2\right)\right]\nonumber\\
&&+\frac{Y_{b1}}{16f}(b_Lb_R^c)^\dag\left[-\frac{2i}{\sqrt{3}}c_\theta h_1\eta_2-(1+c_\theta)(H^-\eta^++(\eta^0)^\ast H^0))-(1-c_\theta)(H^+\eta^-+\eta^0(H^0)^\ast)\right. \nonumber\\
&&\left.\phantom{+\frac{Y_{b1}}{f}(b_Lb_R^c)^\dag[}+s_\theta\left(\eta^+\eta^-+H^+H^-+|H^0|^2+|\eta^0|^2+h_1^2+\eta_1^2+\frac{1}{3}\eta_2^2\right)\right] \nonumber \\
&& + \frac{Y_{t1}}{8 f} (b_L t_R^c)^\dag\left[ \eta^- H^0 - H^- \eta^0 \right] + \frac{Y_{b1}}{8 f} (t_L b_R^c)^\dag\left[\eta^+ (H^0)^\ast - H^+ (\eta^0)^\ast \right]+ \mbox{h.c.}
\end{eqnarray}

\subsection{Self-couplings of the pNGBs} \label{app:pNGBcoupl3}

The potential that determines the misalignment of the vacuum also generates self interactions among the pNGBs, which do not involve derivatives and are proportional to the spurions explicitly breaking the global symmetry in the strong sector. 
Here we will report the results in the vacuum of Section~\ref{sec:i2HDM}, after imposing the minimum condition (and solving for the average techni-fermion mass $M$ as a function of the minimum misalignment angle $\theta$). Furthermore, we define
\begin{equation}
Y_\pm^2\equiv|Y_{t1}|^2\pm|Y_{b1}|^2\,
\end{equation}
and work in the Unitary gauge.

First, the trilinear and quartic Higgs couplings read:
\begin{eqnarray}
\mathcal{L}_{h^3+h^4} &=& \frac{fs_\theta}{96\sqrt{2}}\ \big[ 5C_tY_+^2+\frac{3}{2}C_gg^2(3+t_W^2)\big]\ h_1^3\nonumber\\
&& -\frac{1}{3072}\ \big[2C_tY_+^2(-1+7c_{2\theta})+C_gg^2(3+t_W^2)(1+9c_{2\theta})\big]\ h_1^4\,.
\end{eqnarray}

We also report all the couplings that may be relevant for the calculation of production and scattering rates of the pNGBs (useful for collider studies and Dark Matter ones). The potential generates couplings of a single Higgs field with two other pNGB, in the form
\begin{equation}
\mathcal{L}_{h\phi\phi} = \frac{f}{96 \sqrt{2}}\ g_{h\phi_1\phi_2}\ h_1 \phi_1 \phi_2\,,
\end{equation}
with
\begin{eqnarray}
g_{h_1 H^+ H^-} &=& \big[(14C_tY_+^2+3C_gg^2(3+t_W^2))c_\theta+3C_gg^2(3-3t_W^2)\big] s_\theta\,, \nonumber \\
g_{h_1 (H^0)^\ast H^0} &=& \big[(14C_tY_+^2+3C_gg^2(3+t_W^2))c_\theta+3C_gg^2(3+t_W^2)\big] s_\theta\,, \nonumber\\
g_{h_1 \eta^+ \eta^-} &=& \big[(14C_tY_+^2+3C_gg^2(3+t_W^2))c_\theta-3C_gg^2(3-3t_W^2)\big] s_\theta\,, \nonumber\\
g_{h_1 (\eta^0)^\ast \eta^0} &=& \big[(14C_tY_+^2+3C_gg^2(3+t_W^2))c_\theta-3C_gg^2(3+t_W^2)\big] s_\theta\,, \nonumber\\
g_{h_1 \eta^- H^+} &=& g_{h_1 \eta^+ H^-} = g_{h_1 (\eta^0)^\ast H^0} = g_{h_1 \eta^0 (H^0)^\ast} = 6 C_tY_-^2 c_\theta\,, \nonumber\\
g_{h_1 \eta_1^2} &=& \big[10C_tY_+^2+3C_gg^2(3+t_W^2)\big] c_\theta s_\theta\,, \nonumber\\
g_{h_1 \eta_2^2} &=& \big[2C_tY_+^2-C_gg^2(3+t_W^2)\big] c_\theta s_\theta\,. 
\end{eqnarray}
For completeness, the following terms exhaust the list of trilinear couplings:
\begin{eqnarray}
\mathcal{L}_{\phi^3} &=& \frac{ifs_\theta}{96\sqrt{2}}\ \Big\{(\eta^+H^--\eta^-H^+)\ \big[9C_gg^2(1-t_W^2)\ \eta_1+\sqrt{3}(2C_tY_+^2+C_gg^2(3+t_W^2))c_\theta\ \eta_2 \big]\nonumber\\
&&+( (\eta^0)^\ast H^0-\eta^0(H^0)^\ast) \ \big[3C_gg^2(3+t_W^2)\ \eta_1+\sqrt{3}(2C_tY_+^2+C_gg^2(3+t_W^2))c_\theta\ \eta_2\big] \Big\}\,.
\end{eqnarray}

Quartic couplings that are bilinear in the Higgs field have the form
\begin{equation}
\mathcal{L}_{h_1^2 \phi_1 \phi_2} = - \frac{1}{768} \ g_{h_1^2 \phi_1 \phi_2}\ h_1^2 \phi_1 \phi_2\,,
\end{equation}
with
\begin{eqnarray}
g_{h_1^2 H^+ H^-} &=& \big[-C_tY_+^2(2+c_\theta-7c_{2\theta})+8C_tK_\delta+C_gg^2(3+t_W^2)(1+4c_\theta+5c_{2\theta})+2C_gg^2t_W^2(1-7c_\theta)\big]\,, \nonumber \\
g_{h_1^2 (H^0)^\ast H^0} &=& \big[-C_tY_+^2(2+c_\theta-7c_{2\theta})+8C_tK_\delta+C_gg^2(3+t_W^2)(1+4c_\theta+5c_{2\theta})\big]\,, \nonumber\\
g_{h_1^2 \eta^+ \eta^-} &=& \big[-C_tY_+^2(2+(1-2\Delta)c_\theta-7c_{2\theta})+4C_tK_\delta(1-\Delta)+C_gg^2(3+t_W^2)(1-(3+\Delta)c_\theta+5c_{2\theta}) \nonumber \\
&& +2C_gg^2t_W^2(1+7c_\theta)\big]\,, \nonumber\\
g_{h_1^2 (\eta^0)^\ast \eta^0} &=& \big[-C_tY_+^2(2+(1-2\Delta)c_\theta-7c_{2\theta})+4C_tK_\delta(1-\Delta)+C_gg^2(3+t_W^2)(1-(3+\Delta)c_\theta+5c_{2\theta})\big] \,, \nonumber\\
g_{h_1^2 \eta^- H^+} &=& g_{h_1^2 \eta^+ H^-} = g_{h_1^2 (\eta^0)^\ast H^0} = g_{h_1^2 \eta^0 (H^0)^\ast} = -7 C_tY_-^2 s_\theta\,, \nonumber\\
g_{h_1^2 \eta_1^2} &=& \frac{1}{2} \big[2 C_tY_+^2(-3+5c_{2\theta})+C_gg^2(3+t_W^2)(-1+7c_{2\theta})\big]\,, \nonumber \\
g_{h_1^2 \eta_2^2} &=& \big[-2C_tY_+^2+C_gg^2(3+t_W^2)\big] c_\theta^2\,, \nonumber \\
g_{h_1^2 \eta_1 \eta_2} &=& \frac{2}{\sqrt{3}} \big[-(2C_tY_+^2-C_gg^2(3+t_W^2))\Delta c_\theta+4C_tK_\delta(1+\Delta)\big]\,. 
\end{eqnarray}
Quartic couplings linear in the Higgs are also generated in the form
\begin{eqnarray}
\mathcal{L}_{h_1 \phi^3} &=& \frac{i}{768} h_1\ \big[ (H^0)^\ast\eta^0-H^0(\eta^0)^\ast \big]\  \big[  g_{h_1 \eta_1 00}\ \eta_1 + g_{h_1 \eta_2 00}\ \eta_2 \big] + \nonumber \\
&& \frac{i}{768} h_1\ \big[ H^+\eta^--H^-\eta^+ \big]\  \big[  g_{h_1 \eta_1 +-}\ \eta_1 + g_{h_1 \eta_2 +-}\ \eta_2 \big]\,,
\end{eqnarray}
with
\begin{eqnarray}
g_{h_1 \eta_1 00} &=& \big[ (2C_tY_+^2\Delta+C_gg^2(3+t_W^2)(3-\Delta))c_\theta-4C_tK_\delta(1+\Delta)\big]\,, \nonumber \\
g_{h_1 \eta_2 00} &=& \frac{1}{\sqrt{3}} \big[ 2C_tY_+^2(2-(1-\Delta)c_\theta+5c_{2\theta})-4C_tK_\delta(-3+\Delta) \nonumber \\
&& \phantom{aaaaaaaaaa} -C_gg^2(3+t_W^2)(2-(1-\Delta)c_\theta-4c_{2\theta}) \big]\,, \nonumber \\
g_{h_1 \eta_1 +-} &=& \big[ (2C_tY_+^2\Delta+C_gg^2(3+t_W^2)(3-\Delta))c_\theta-4C_tK_\delta(1+\Delta)+12C_gg^2t_W^2c_\theta\big]\,, \nonumber \\
g_{h_1 \eta_2 +-} &=&  \frac{1}{\sqrt{3}}  \big[ 2C_tY_+^2(2-(1-\Delta)c_\theta+5c_{2\theta})-4C_tK_\delta(-3+\Delta)\nonumber\\
&&\phantom{aaaaaaaaaa}-C_gg^2(3+t_W^2)(2-(1-\Delta)c_\theta-4c_{2\theta})+12C_gg^2t_W^2\big]\,. 
\end{eqnarray}

%%%%%%%%%%%%%%%%%%%%%%%%%%%%%%%%%%%%%%%%%
\section{General vacuum}

\subsection{Parameterisation of the vacuum} \label{app:genvac}

The form of the most general vacuum misalignment can be simplified by using the following property of the $\SU(6)$ generators.\\

{\it Consider three generators $\{ A, B, C\}$ of $\SU(N)$ (i.e. $N\times N$ hermitian matrices) that form an $\SU(2)$ algebra 
\begin{equation}
\big[ A, B\big] = \frac{i}{\kappa} C\,, \qquad \big[B, C\big] = \frac{i}{\kappa} A\,, \qquad \big[C, A\big] = \frac{i}{\kappa} B\,.
\end{equation}
Then, the following relation holds:
\begin{equation} \label{eq:rule}
e^{i (a A + b B)} = e^{- i \kappa \varphi\ C} \cdot e^{i \rho\ A} \cdot  e^{i \kappa \varphi\ C}\,,
\end{equation}
where $a = \rho \cos \varphi$ and $b = \rho \sin \varphi$. }\\

The most general CP-conserving vacuum, misaligned along the three directions identified in Section~\ref{sec:model}, is defined by the rotation
\begin{equation}
\Omega = e^{i \sqrt{2} (\theta_1 X^4 + \theta_2 X^8 + \theta_3 X^{13})}\,.
\end{equation}
We remark that the following triplets of generators form (different) $\SU(2)$ subgroups of $\SU(6)$
\begin{equation}
\{ X^4,\ X^{13},\ S^{14} \}\,, \qquad \{ X^{13},\ X^8,\ S^{10} \}\,,
\end{equation}
with $\kappa = 2 \sqrt{2}$.
Thus, applying the relation in Eq.(\ref{eq:rule}), the rotation can be written as
\beq
\Omega = e^{- i 2 \sqrt{2} \gamma (c_\beta S^{14} - s_\beta S^{10})} \cdot e^{i \sqrt{2} \theta (c_\beta X^4 + s_\beta X^8)} \cdot e^{i 2 \sqrt{2} \gamma (c_\beta S^{14} - s_\beta S^{10})}\,,
\eeq
where $\theta_1 = \theta c_\gamma s_\beta$, $\theta_2 = \theta c_\gamma s_\beta$ and $\theta_3 = \theta s_\gamma$.
The relation in Eq.(\ref{eq:rule}) can be applied again to the two exponentials, as the following triplets also form $\SU(2)$ subgroups of $\SU(6)$ (with the same value of $\kappa$):
\begin{equation}
\{ X^8,\ X^{4},\ S^{21} \}\,, \qquad \{ S^{14},\ S^{10},\ S^{21} \}\,.
\end{equation}
Thus:
\begin{eqnarray}
e^{i \sqrt{2} \theta (c_\beta X^4 + s_\beta X^8)} &=& e^{i 2 \sqrt{2} \beta S^{21}} \cdot e^{i \sqrt{2} \theta X^4}\cdot e^{-i 2 \sqrt{2} \beta S^{21}} \,, \\
e^{- i 2 \sqrt{2} \gamma (c_\beta S^{14} - s_\beta S^{10})} &=& e^{i 2 \sqrt{2} \beta S^{21}} \cdot e^{- i 2 \sqrt{2} \gamma S^{14}} \cdot  e^{-i 2 \sqrt{2} \beta S^{21}}\,.
\end{eqnarray}
(The former relation explains Eq.(\ref{eq:Omegatb}).) Putting these relations together allows us to write
\begin{equation}
\Omega = e^{i 2 \sqrt{2} \beta S^{21}} \cdot e^{- i 2 \sqrt{2} \gamma S^{14}} \cdot e^{i \sqrt{2} \theta X^4} \cdot e^{i 2 \sqrt{2} \gamma S^{14}} \cdot  e^{-i 2 \sqrt{2} \beta S^{21}}\,,
\end{equation}
which coincides with the one used in Eq.(\ref{eq:Omegatbg}) of Section~\ref{sec:vacuum}.

\subsection{Masses of the pNGBs} \label{app:pNGB}

The Goldstones eaten by the massive $W$ and $Z$ are defined as
\begin{align}
\pi'_3 & = \frac{1}{\cos \frac{\tau}{2}} \left( \cos \gamma\ \cos \frac{\theta}{2} \ \pi_3 + \sin \gamma\ \eta_1 \right)\,, \\
{\pi'}^\pm & = \frac{1}{\cos \frac{\tau}{2}} \left(\cos \gamma\ \cos \frac{\theta}{2} \ \frac{\pi_1 \pm i \pi_2}{\sqrt{2}} + \sin \gamma\  \eta^\pm \right)\,,
\end{align}
while the orthogonal combinations remain as physical states:
\begin{align}
\eta'_1 & = \frac{1}{\cos \frac{\tau}{2}} \left(\cos \gamma \cos \frac{\theta}{2} \ \eta_1 - \sin \gamma\ \pi_3 \right)\,, \\
{\eta'}^\pm & = \frac{1}{\cos \frac{\tau}{2}} \left(\cos \gamma\ \cos \frac{\theta}{2} \ \eta^\pm  + \sin \gamma\  \frac{\pi_1 \pm i \pi_2}{\sqrt{2}} \right)\,.
\end{align}

The mass matrix for the 3 scalars $h_1'$, $h_2$ and $\varphi_0'$ at the minimum of the vacuum as defined in Section~\ref{sec:minimiz} can be written, in matrix form, as
\begin{equation}
\frac{1}{2} \frac{C_t f^2}{4} \left(h_1'\ , \; h_2\ , \; \varphi_0' \right) \cdot N \cdot \left( \begin{array}{c}
h_1' \\ h_2 \\ \varphi_0' \end{array} \right)\,,
\end{equation}
with the symmetric matrix $N$ having entries:
\begin{eqnarray}
N_{11} &=&  \left(Y_t^2 - \frac{C_g}{C_t} \frac{2 g^2 + {g'}^2}{2}\right) s^2_\tau\,, \\
N_{12} = N_{21} &=& 4 K_\delta s_{2\beta}\ \frac{s_\frac{\tau}{2}^2}{c_\frac{\tau}{2}}\,, \\
N_{13}=N_{31} &=& - 4 K_\delta s_{2\beta}\ \frac{c_\tau s_\frac{\tau}{2}}{c_\frac{\tau}{2}^2}\,, \\
N_{22} &=& Y_t^2 c_\frac{\tau}{2}^2 - 4 K_\delta c_{2\beta} - 4 \frac{K_\delta^2 s_{2\beta}^2}{Y_t^2}\ \frac{1}{c_\frac{\tau}{2}^2} \,, \\
N_{23} = N_{32} &=& \frac{Y_t^2}{2} s_\tau - 2 \frac{K_\delta^2 s_{2\beta}^2}{Y_t^2}\ \frac{c_\tau}{c_\frac{\tau}{2}^4}\,, \\
N_{33} &=& Y_t^2 s_\frac{\tau}{2}^2 + K_\delta s_{2\beta} \frac{3 c_\tau-1+c_{2\gamma} (3-c_\tau)}{\sqrt{2(c_{2\gamma} + c_\tau)}} \frac{s_\frac{\tau}{2}}{c_\frac{\tau}{2}^2} - 4 \frac{K_\delta^2 s_{2\beta}^2}{Y_t^2}\ \frac{s_\frac{\tau}{2}^2}{c_\frac{\tau}{2}^4}\,.
\end{eqnarray}
Note that in absence of mixing, i.e. $K_\delta \ll 1$, the mass of $h_1'$,
\begin{equation}
m_{h_1'}^2 = \frac{C_t f^2}{4} N_{11}= C_t \frac{m_t^2}{4} - C_g \frac{2 m_W^2 + m_Z^2}{16}\,,
\end{equation}
matches the result of the minimal $\SU(4)/\Sp(4)$ model.

\bibliography{SU6litt.bib}

\end{document}